\documentclass[prd,preprint,floatfix,a4paper,
showpacs,showkeys,superscriptaddress,nofootinbib]{revtex4}%

\usepackage{graphicx}
\graphicspath{./}

\usepackage{amssymb,amsmath,amsfonts,amsthm}
\usepackage{array,multirow}

\usepackage[colorlinks=true,linkcolor=blue,citecolor=blue]{hyperref}
\usepackage{url}  
\newcommand{\Fig}[1]{Fig.~\ref{#1}}
\newcommand{\Tab}[1]{Table~\ref{#1}}
\newcommand{\Sec}[1]{Sec.~\ref{#1}}
\newcommand{\Secs}[1]{Secs.~\ref{#1}}
\newcommand{\App}[1]{Appendix~\ref{#1}}
\newcommand{\Eq}[1]{Eq.\,\eqref{#1}}
\newcommand{\Ref}[1]{Ref.\,\cite{#1}}
\newcommand{\Refs}[1]{Refs.\,\cite{#1}}

\newcommand{\Tr}{\operatorname{Tr}}

\renewcommand{\Re}{\operatorname{\mathfrak{Re}}}
\providecommand{\abs}[1]{\lvert#1\rvert}

\begin{document}


\title{Scaling study of the gluon propagator in Coulomb gauge QCD
on isotropic and anisotropic lattices}


\author{Y.~Nakagawa}
\email{nakagawa@muse.sc.niigata-u.ac.jp}
\affiliation{Graduate School of Science and Technology,
Niigata University, Niigata 950-2181, Japan}

\author{A.~Nakamura}
\affiliation{Research Institute for Information Science and Education,
Hiroshima University, Hiroshima 739-8521, Japan}

\author{T.~Saito}
\affiliation{Integrated Information Center, 
Kochi University, Kochi 780-8520, Japan}

\author{H.~Toki}
\affiliation{Research Center for Nuclear Physics, 
Osaka University, Osaka 567-0047, Japan}


\begin{abstract}
We calculate the transverse and time-time components of the
instantaneous gluon propagator in Coulomb gauge QCD by using
an SU(3) quenched lattice simulation on isotropic and
anisotropic lattices.
We find that the gluon propagators suffer from strong discretization
effects on the isotropic lattice; on the other hand, those on
the anisotropic lattices give a better scaling.
Moreover, on these two type of lattices the transverse parts
are significantly suppressed in the infrared region and
have a turnover at about 500 [MeV].
The high resolution to the temporal direction due to the anisotropy
yields small discretization errors for the time-time gluon propagators,
which also show an infrared enhancement
as expected in the Gribov-Zwanziger confinement scenario.
\end{abstract}


\pacs{11.15.Ha, 12.38.Gc, 12.38.Aw}
\keywords{lattice QCD, color confinement, Coulomb gauge, gluon propagator}


\maketitle


\section{Introduction}
\label{sec:Introduction}


The Coulomb gauge with no negative metric provides a very clear physical
picture in the sense that the color-Gauss's law can be formally solved,
only transverse degrees of freedom appear as dynamical degrees of freedom,
and Fock space is well defined.
The prominent feature taking the Coulomb gauge is that
an instantaneous interaction, which is requisite for color confinement,
shows up in the Hamiltonian. In the Gribov-Zwanziger scenario,
the path integral is dominated by the configurations near the Gribov
horizon where the lowest eigenvalue of the Faddeev-Popov (FP) ghost
operator vanishes
\cite{Zwanziger:1993dh}.
The lattice simulations show the
enhancement of near-zero modes of the FP eigenvalues
\cite{Greensite:2004ur,Nakagawa:2007fa}.
Accordingly, the color-Coulomb instantaneous interaction bears
a confining force.
It has been also confirmed by the lattice simulations that
the color-Coulomb potential rises linearly at large distances and
its string tension is larger than the string tension of the static Wilson potential
\cite{Greensite:2003xf,Nakamura:2005ux,
Nakagawa:2006fk,Nakagawa:2008ip},
which is expected from the Zwanziger's inequality
\cite{Zwanziger:2002sh}.
In addition, the color-Coulomb potential can be reevaluated by
inverting the FP ghost matrix and this analysis has shown that
its string tension almost saturates the Wilson string tension
\cite{Voigt:2008rr}.


Exploring the gluon propagator is a central issue
for studying the confinement mechanism in QCD.
The transverse would-be physical gluon propagator is expected to be suppressed
in the infrared (IR) region due to the proximity of the Gribov region
in the IR direction in the Gribov-Zwanziger scenario
\cite{Zwanziger:1991gz}.
There have been a lot of lattice studies and functional analyses
in the Landau gauge
(see, for instance, \Ref{Bogolubsky:2009dc,Fischer:2008uz}
and references therein).
Contrastingly, there are few lattice studies on
the instantaneous gluon propagator in the Coulomb gauge
\cite{Cucchieri:2000gu,Langfeld:2004qs,
Burgio:2008jr,Nakagawa:2009zf}.
In the continuum theory, on the other hand, there are many works on
the variational approach in the Coulomb gauge, which are very useful
to study the spectrum of the hadronic bound states and the Green's functions
\cite{Szczepaniak:2001rg,Szczepaniak:2003ve,Feuchter:2004mk,
Epple:2006hv,Epple:2007ut,Campagnari:2008yg},
in addition to the functional analysis
\cite{Watson:2007vc,Reinhardt:2008pr,Watson:2009fb}.

The Coulomb gauge fixing condition is imposed on each time slice
and it does not introduce correlations between neighboring time slices.
Accordingly, the equal-time correlation between gauge fields
at different points
\begin{equation}
\langle A^a_{\mu}(\vec{x},t) A^b_{\nu}(\vec{y},t) \rangle
\end{equation}
is well defined in the Coulomb gauge.
Recent lattice studies of the instantaneous gluon propagator
revealed that it shows scaling violation
\cite{Burgio:2008jr,Nakagawa:2009zf}; namely,
the gluon propagator calculated at different lattice couplings
do not fall on top of one curve after multiplicative renormalization.


In order to circumvent the problem of scaling violation,
the authors of \Ref{Burgio:2008jr}
have extracted the instantaneous gluon propagator by eliminating
the $p_4$ dependence of the unequal-time propagator.
It has been concluded that the instantaneous transverse gluon propagator
$D^{\rm tr}(\vec{p})$ is multiplicatively renormalizable
in the Hamiltonian limit and the numerical data of it are well fitted
with the Gribov-type form of the propagator
\footnote{
In this paper, we use the same symbol for the instantaneous propagator
and the unequal-time propagator, but the reader may distinguish them
by their argument: the instantaneous propagator does not depend on
$x_4$ (or $p_4$ in momentum space) but the unequal-time one does.
}
\begin{equation}\label{eq:GribovAnsatz}
D^{\rm tr}(\vec{p})
= \frac{1}{2\sqrt{\abs{\vec{p}}^2+\frac{M^2}{\abs{\vec{p}}^2}}},
\end{equation}
where $M$ is a fitting parameter at which the propagator shows a turnover.
The method was applied to the transverse gluon propagator successfully
both in 2+1 and 3+1 dimensional SU(2) Yang-Mills theory
\cite{Burgio:2009xp,Burgio:2008jr}
while it does not improve the scaling violation of the temporal gluon propagator
since the time-time component of the unequal-time gluon propagator is
energy independent even after the residual gauge fixing
\cite{Quandt:2008zj}.


As another approach, a new momentum cut is introduced in
\Ref{Nakagawa:2009zf}
in addition to the cone cut and the cylinder cut.
High momentum data that suffer from discretization errors are excluded
from the analysis of the instantaneous propagators by this new cut.
Combined with the matching analysis given in
\Ref{Leinweber:1998uu},
it has been shown that this procedure successfully reduces
the scaling violation for the transverse gluon propagator,
but it fails for the time-time component of the gluon propagator.


In this study, we take a different route to show that
the instantaneous gluon propagator is multiplicatively renormalizable.
The problem of scaling violation of the instantaneous propagator
can be seen even at the tree level with a finite temporal lattice spacing
as was discussed in
\Ref{Burgio:2008jr}
(we shall briefly review the point in \App{sec:Dtr_free}).
The reason is that the energy integral does not run from $-\infty$
to $\infty$ but from $-\pi/a_{\tau}$ to $\pi/a_{\tau}$ on a finite lattice,
and this introduces the spurious $\abs{\vec{p}}$ dependence on
the free instantaneous propagator.
Therefore, we expect that the instantaneous propagator is
multiplicatively renormalizable in the Hamiltonian limit
$a_{\tau} \to 0$.
To make this point clear, we calculate the transverse and
the time-time components of the instantaneous gluon propagator
on anisotropic lattices and show how the scaling violation becomes milder
as the anisotropy increases, i.e., as we get close to the Hamiltonian limit.


The organization of this paper is as follows.
In the subsequent sections, we describe the lattice observables,
the space and time components of the instantaneous gluon propagator,
and the lattice setup of our numerical simulations.
In \Secs{sec:Dtr_isotropic} and \ref{sec:D44_isotropic},
the numerical results of the transverse and the temporal components
of the instantaneous propagators on the isotropic lattice are reported.
Section \ref{sec:Dtr_anisotropic} is devoted to show the results for
the transverse propagator on the anisotropic lattices.
In the subsequent \Secs{sec:Dtr_anisotropic_fit},
we discuss the IR and the ultraviolet (UV) behavior
of the propagator by making the power law fitting.
The anisotropic lattice results for the temporal gluon propagator
are given in \Sec{sec:Z44_anisotropic},
and the IR and the UV fittings are examined in \Sec{sec:Z44_anisotropic_fit}.
We attempt to extract the color-Coulomb string tension
from the instantaneous temporal gluon propagator in position space
in \Sec{sec:D_44_and_Vc}.
The conclusions are drawn in \Sec{sec:conclusions}.

\section{Instantaneous gluon propagator}
\label{sec:observables}


We calculate the transverse and the time-time components
of the instantaneous gluon propagator,
\begin{equation}
D^{ab}_{\mu\nu}(\vec{x}-\vec{y})
= \langle A^a_{\mu}(\vec{x},t)A^b_{\nu}(\vec{y},t)\rangle,
\end{equation}
in the momentum space,
\begin{align}
& D^{ab}_{ij}(\vec{p})
  = \delta^{ab}\left(\delta_{ij}-\frac{p_ip_j}{\abs{\vec{p}}^2}\right)
  D^{\rm tr}(\vec{p}), \\
& D^{ab}_{44}(\vec{p})
  =\delta^{ab}\frac{Z_{44}(\vec{p})}{\abs{\vec{p}}^2},
\end{align}
where the gauge fields are related to the link variables through
\begin{equation}\label{eq:linear_def}
A^{\rm lat}_{\mu}(\vec{x},t)
= \left.
\frac{U_{\mu}(\vec{x},t)-U_{\mu}^{\dagger}(\vec{x},t)}{2ig_0a_{\mu}}
\right|_{\rm traceless}. 
\end{equation}
The instantaneous gluon propagator is evaluated on each time slice
and we average over all time slices.
The lattice momenta $k_{\mu}$ are discretized
and take integer values in the range $(-L_{\mu}/2, L_{\mu}/2]$.
The lattice momenta and the continuum ones are related via
\begin{equation}
p_{\mu} = \frac{2}{a_{\mu}} \sin \left( \frac{\pi k_{\mu}}{L_{\mu}} \right),
\end{equation}
where $a_{\mu} (L_{\mu})$ are $a_{\sigma} (L_{\sigma})$ for
$\mu = 1$ to $3$ and $a_{\tau} (L_{\tau})$ for $\mu=4$, respectively.

We note that the unequal-time propagator $D_{\mu\nu}$ has
mass dimension 2 in momentum space, while the instantaneous one
has mass dimension 1.
This is because the instantaneous propagator is obtained by integrating
the unequal-time propagator over the time component of the four momentum;
\begin{equation}
D_{\mu\nu}(\vec{p}) = \int \frac{dp_4}{2\pi} D_{\mu\nu}(\vec{p},p_4).
\end{equation}

The unrenormalized transverse gluon propagator $D^{\rm tr}_{\rm lat}$,
which is measured by lattice simulations,
is related to the renormalized propagator $D^{\rm tr}_{\rm R}$
via the multiplicative renormalization,
\begin{equation}
D^{\rm tr}_{\rm R}(\vec{p};\mu) = a_{\sigma} Z^{\rm tr}(a_{\sigma},\mu) D^{\rm tr}_{\rm lat}(\vec{p}a_{\sigma}),
\end{equation}
where $\mu$ is the renormalization point.
We expect that the renormalized propagator is independent of
the lattice spacing in the scaling regime.
As we shall see later, the multiplicative renormalizability
does not hold for finite temporal lattice spacing
and we have to take the Hamiltonian limit, $a_{\tau} \to 0$.


The color-Coulomb potential plays a crucial role in the Coulomb gauge QCD.
It was shown that the time-time component of the gluon propagator
can be decomposed into the instantaneous part and the noninstantaneous part
\cite{Cucchieri:2000hv},
\begin{equation}\label{eq:uneqD44}
D_{44}(\vec{x},t) = V_c(\vec{x})\delta(t) + P(\vec{x},t).
\end{equation}
The first term in the right-hand side represents the color-Coulomb
potential, which is defined as the vacuum expectation value of
the kernel of the instantaneous interaction,
\begin{equation}
V_c(\vec{x}-\vec{y}) \delta^{ab}
= \langle (M^{-1}[A](-\partial_i^2)M^{-1}[A])^{ab}_{\vec{x},\vec{y}} \rangle.
\end{equation}
Here $M^{-1}$ is the Green's function of the Faddeev-Popov ghost operator.
In \Eq{eq:uneqD44}, $P(\vec{x},t)$ is assumed to be nonsingular at $t=0$
as opposed to the first term.
It has been shown that both $V_c$ and $P$ are
renormalization-group invariant
\cite{Zwanziger:1998ez};
namely, the renormalization constant
for the color-Coulomb potential can be set to be 1 in the continuum limit.
We expect that the color-Coulomb potential $V_c$ can be extracted
from the instantaneous temporal gluon propagator as
\begin{equation}
V_c(\vec{x}) = a_{\tau} D_{44}(\vec{x}),
\end{equation}
when we are in the scaling region and the lattice spacing
is small enough, where $a_{\tau}$ comes from the $\delta$ function.

\section{Lattice setup}
\label{sec:lattice_setup}


The lattice configurations are generated by
the heat-bath Monte Carlo technique with
the standard Wilson plaquette action,
\begin{equation}
  S = \frac{\beta}{\xi_B} \sum_{n, i < j \le 3}
  \Re\Tr (1-U_{ij}(n)) \\
    + \beta \xi_B \sum_{n, i \le 3}
  \Re\Tr (1-U_{i4}(n)).
\end{equation}
Here $U_{\mu\nu}(n)$ indicates the plaquette operator, and
$\beta=2 N_c / g_0^2$ is the lattice coupling.
On the isotropic lattice, the bare anisotropy $\xi_B$ is 1
and the action can be written in a familiar form
\begin{equation}
  S = \beta \sum_{n, \mu < \nu} \Re\Tr (1-U_{\mu\nu}(n)). \\
\end{equation}
The renormalized anisotropy $\xi$ is defined as the ratio of
the spatial lattice spacing to the temporal lattice spacing,
$\xi=a_{\sigma}/a_{\tau}$.
The ratio of $\xi_B$ and $\xi$ can be determined nonperturbatively by
matching the spatial and the temporal Wilson loops on anisotropic lattices.
We use the relation obtained by Klassen for the range $1 \le \xi_B \le 6$
and $5.5 \le \beta \le \infty$
\cite{Klassen:1998ua}.
We adopt the values of the lattice spacing given in
\Ref{Namekawa:2001ih}
for $\xi=2$ and in
\Ref{Matsufuru:2001cp}
for $\xi=4$,
where the static quark potential was measured
to set the scale.
For the isotropic lattice, the scale is set by using
the scaling relation obtained by Necco and Sommer,
with the Sommer scale parameter $r_0 = 0.5$ [fm],
which is applicable in the range $5.7 \le \beta \le 6.92$
\cite{Necco:2001xg}.


In our simulations,
the first 5000 sweeps are discarded for thermalization,
and we measured the instantaneous gluon propagator for $50-100$ configurations,
each of which is separated by 100 sweeps.
All the lattice parameters are given in \Tab{tab:lattice_setup}.

\begin{table*}[htbp]
\caption{
The lattice couplings, the spatial and the temporal lattice extents,
the lattice spacings, the lattice volumes in physical units,
and the number of configurations used to evaluate the instantaneous propagators.
}
\begin{center}\begin{tabular}{ccccccccc}
\hline\hline 
$\xi=a_{\sigma}/a_{\tau}$ & $L_{\sigma}$ & $L_{\tau}$ & $\beta$ & $\xi_B$
& $a_{\sigma}^{-1}$ [GeV] & $a_{\sigma}$ [fm] & $V$[fm$^4$] & Number of configurations \\
\hline \hline
\multirow{10}{*}{1}
 & 24 & 24 & 5.70 & 1     & 1.160 & 0.1702 & 4.09$^4$ & 100 \\
 & 48 & 48 &  :   &   :   &   :   &   :    & 8.17$^4$ & 100 \\
 & 56 & 56 &  :   &   :   &   :   &   :    & 9.53$^4$ & 100 \\
 & 64 & 64 &  :   &   :   &   :   &   :    & 10.9$^4$ & 100 \\
 & 24 & 24 & 5.80 &   :   & 1.446 & 0.1364 & 3.27$^4$ & 100 \\
 & 24 & 24 & 6.00 &   :   & 2.118 & 0.0932 & 2.24$^4$ & 100 \\
 & 32 & 32 &  :   &   :   &   :   &   :    & 2.98$^4$ & 100 \\
 & 48 & 48 &  :   &   :   &   :   &   :    & 4.47$^4$ & 100 \\
 & 56 & 56 &  :   &   :   &   :   &   :    & 5.22$^4$ & 100 \\
 & 64 & 64 &  :   &   :   &   :   &   :    & 5.97$^4$ & 100 \\
\hline
\multirow{3}{*}{2}
 & 24 & 48 & 5.80 & 1.674 & 1.104 & 0.1787 & 4.29$^4$ & 80 \\
 & 24 & 48 & 6.00 & 1.705 & 1.609 & 0.1227 & 2.94$^4$ & 80 \\
 & 24 & 48 & 6.10 & 1.718 & 1.889 & 0.1045 & 2.51$^4$ & 80 \\
\hline
\multirow{9}{*}{4}
 & 16 &  64 & 5.75 & 3.072 & 1.100 & 0.1794 & 2.87$^4$ & 100 \\
 & 24 &  96 &  :   &   :   &   :   &    :   & 4.31$^4$ & 50 \\
 & 32 & 128 &  :   &   :   &   :   &    :   & 5.74$^4$ & 50 \\
 & 48 & 192 &  :   &   :   &   :   &    :   & 8.61$^4$ & 100 \\
 & 24 &  96 & 5.95 & 3.159 & 1.623 & 0.1216 & 2.92$^4$ & 50 \\
 & 48 & 192 &  :   &   :   &   :   &    :   & 5.84$^4$ & 100 \\
 & 24 &  96 & 6.10 & 3.211 & 2.030 & 0.0972 & 2.33$^4$ & 50 \\
 & 32 & 128 &  :   &   :   &   :   &    :   & 3.11$^4$ & 50 \\
 & 48 & 192 &  :   &   :   &   :   &    :   & 4.67$^4$ & 100 \\
\hline\hline
\end{tabular}
\label{tab:lattice_setup}
\end{center}
\end{table*}


In the Coulomb gauge the transversality condition
\begin{equation}
\partial_i A_i(\vec{x},t) = 0
\end{equation}
is imposed on the gauge fields at each time slice,
where $i$ runs from 1 to 3.
On a lattice, gauge configurations satisfying the Coulomb gauge
condition can be obtained by minimizing the functional
\begin{equation}\label{F_U}
F_U[g] = \frac{1}{L_{\sigma}^3} \sum_{\vec{x},i} \Re\Tr
\left( 1 - \frac{1}{3}g^{\dagger}(\vec{x},t)
U_i(\vec{x},t)g(\vec{x}+\vec{i},t) \right)
\end{equation}
with respect to the gauge transformation $g(\vec{x},t) \in$ SU(3)
on each time slice.
The functional derivative of \Eq{F_U} with respect to $g$
yields the transversality condition $\nabla_i A^{\rm lat}_i(\vec{x},t)=0$,
where $\nabla$ is the lattice backward difference, and
it reproduces the Coulomb gauge condition in the continuum limit.
The Coulomb gauge fixing has been done using an iterative method
with the Fourier acceleration
\cite{Davies:1987vs},
and we stop the iterative gauge fixing
if the violation of the transversality becomes less than $10^{-14}$;
\begin{equation}
\theta = \frac{1}{(N_c^2-1)L_{\sigma}^3}\sum_{\vec{x}, a, i}
(\nabla_i A_i^{\rm lat}(\vec{x},t))^2 < 10^{-14}.
\end{equation}
This stopping criterion is applied for each time slice.
We note that the accuracy of the gauge fixing is crucial
for the transverse propagator to see the IR suppression,
which is discussed in \App{sec:tolerance}.


In order to reduce lattice artifacts, we apply
the cone and cylinder cuts to the momenta
\cite{Leinweber:1998uu}.
The cone cut is necessary to address finite volume effects
that are seen in small momentum data.
On the other hand, the cylinder cut reduces artifacts
due to the broken rotational symmetry on lattice.
The statistical errors are estimated by the jackknife method.

\section{Instantaneous transverse gluon propagator on the isotropic lattice}
\label{sec:Dtr_isotropic}

\begin{figure}
\begin{center}
\resizebox{0.45\textwidth}{!}
{\includegraphics{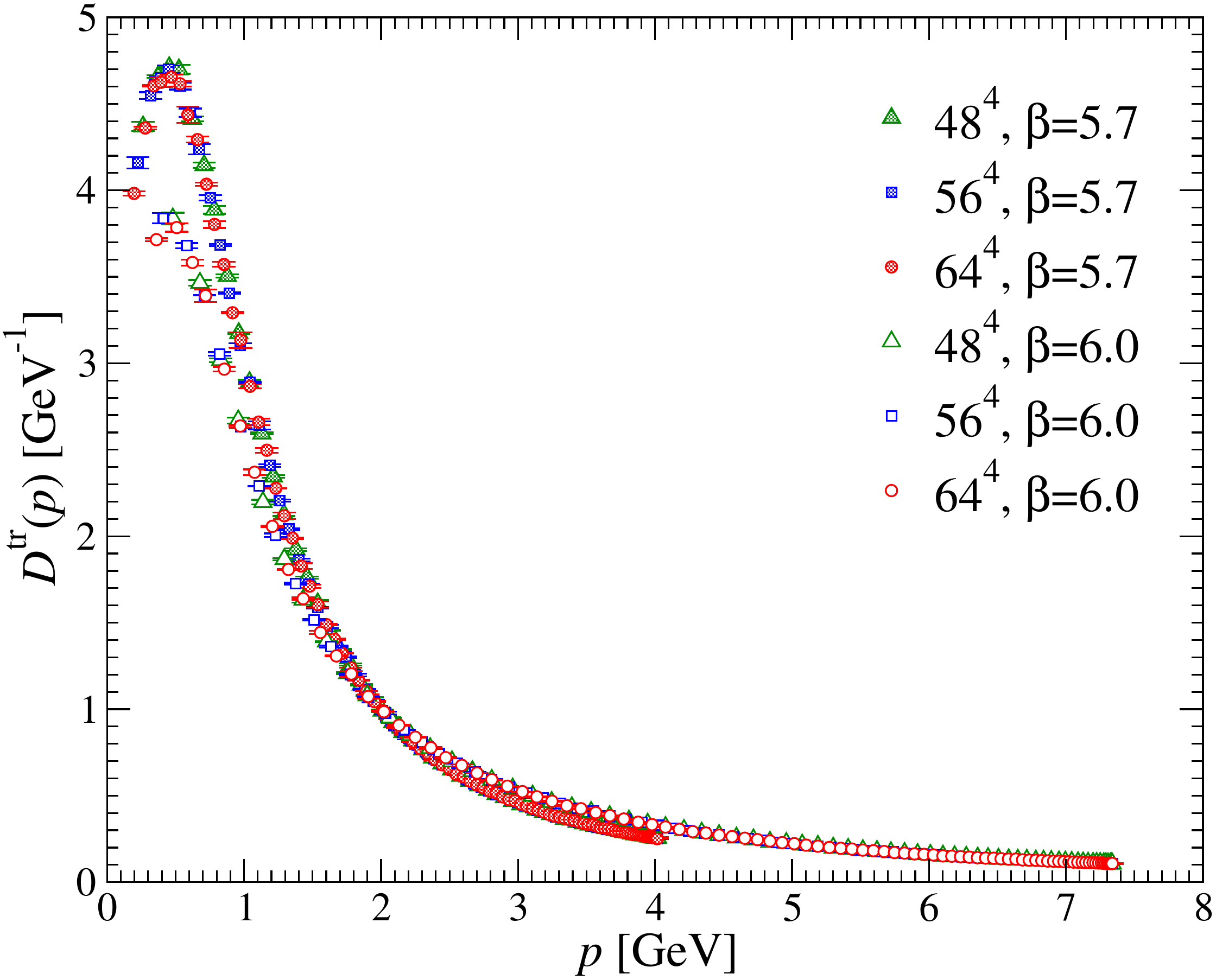}}
\end{center}
\caption{
The instantaneous transverse gluon propagator
in physical units at $\beta=5.7$ and $\beta=6.0$.
The propagator is renormalized at $\abs{\vec{p}}$=2 [GeV].
}
\label{fig:Dtr_large_volume}
\end{figure}

Although the problem of the scaling violation for
the instantaneous gluon propagator has already been discussed in
\Refs{Burgio:2008jr,Nakagawa:2009zf},
we here show the lattice result for the instantaneous
transverse gluon propagator on the isotropic lattice
to clarify the issues.
The instantaneous transverse gluon propagator
on the isotropic lattice at $\beta$=5.7 and 6.0 is drawn
in \Fig{fig:Dtr_large_volume}.
The propagator is normalized such that
$D^{\rm tr}(\abs{\vec{p}}=2{\rm [GeV]}) = 1$.


We observe that $D^{\rm tr}$ has a maximum
at $p = 0.4-0.5$ [GeV] irrespective of the lattice coupling
and it decreases with the momentum in the IR region.
This is a striking feature of the transverse gluon propagator.
The instantaneous propagator is defined as the energy integral
of the unequal-time propagator,
\begin{equation}
D(\abs{\vec{p}}) = \int \frac{dp_4}{2\pi} D(\vec{p}, p_4).
\end{equation}
For a massless particle and a massive particle,
$D(\abs{\vec{p}})=1/(2\abs{\vec{p}})$ and
$D(\abs{\vec{p}}) =1/(2\sqrt{\abs{\vec{p}}^2+m^2})$, respectively.
Thus the instantaneous transverse propagator can be interpreted
as the inverse of the energy dispersion relation of
the would-be physical gluons.
It implies that the propagator at vanishing momentum corresponds to
the inverse of the effective mass of the gluon.
The IR suppression of the instantaneous transverse gluon propagator
means that the gluons have momentum dependent effective mass
$M(\vec{p})$ and it diverges in the IR limit,
$\lim_{\vec{p} \to 0} M(\vec{p})=\infty$,
indicating the confinement of gluons.

In addition to the bump structure of the transverse propagator,
the two curves corresponding to different lattice couplings
cross at the renormalization point and deviate from each other
in the small and the large momentum regions.
This is not seen in the Landau gauge gluon propagator
(see e.g. \Ref{Leinweber:1998uu}).
Such a behavior of the propagator casts doubt the validity
of the multiplicative renormalizability for the instantaneous gluon propagator.


Taking a closer look at the raw results of the numerical simulations
gives us a clue to cure scaling violation.
Assuming multiplicative renormalization for the propagator,
the renormalized dressing function $Z^{\rm tr}_{\rm R}$ of the transverse
gluon propagator, $Z^{\rm tr}_{\rm R}=\abs{\vec{p}}D^{\rm tr}_{\rm R}$,
is related to the bare unrenormalized dressing function $Z^{\rm tr}_{\rm lat}$ via 
\begin{equation}
Z^{\rm tr}_{\rm R}(\vec{p};\mu) = Z(a_{\sigma},\mu) Z^{\rm tr}_{\rm lat}(\abs{\vec{p}}a_\sigma)
\end{equation}
in the scaling regime.
Here $Z(a_{\sigma},\mu)$ is a renormalization constant.
It is easily read off from this relation that
in the log-log plot of the dressing function of the propagator,
converting from lattice units to physical units corresponds to
the parallel displacement in the horizontal direction
and the renormalization of $Z^{\rm tr}$ corresponds to
that in the vertical direction.
If the propagator is multiplicatively renormalizable,
the different curves associated with the different lattice couplings
can fall on top of each other by a parallel shift of the curves
in the horizontal and the vertical directions in a double-log plot.

\begin{figure*}[htbp]
\begin{minipage}{0.45\hsize}\begin{center}
\resizebox{1.\textwidth}{!}
{\includegraphics{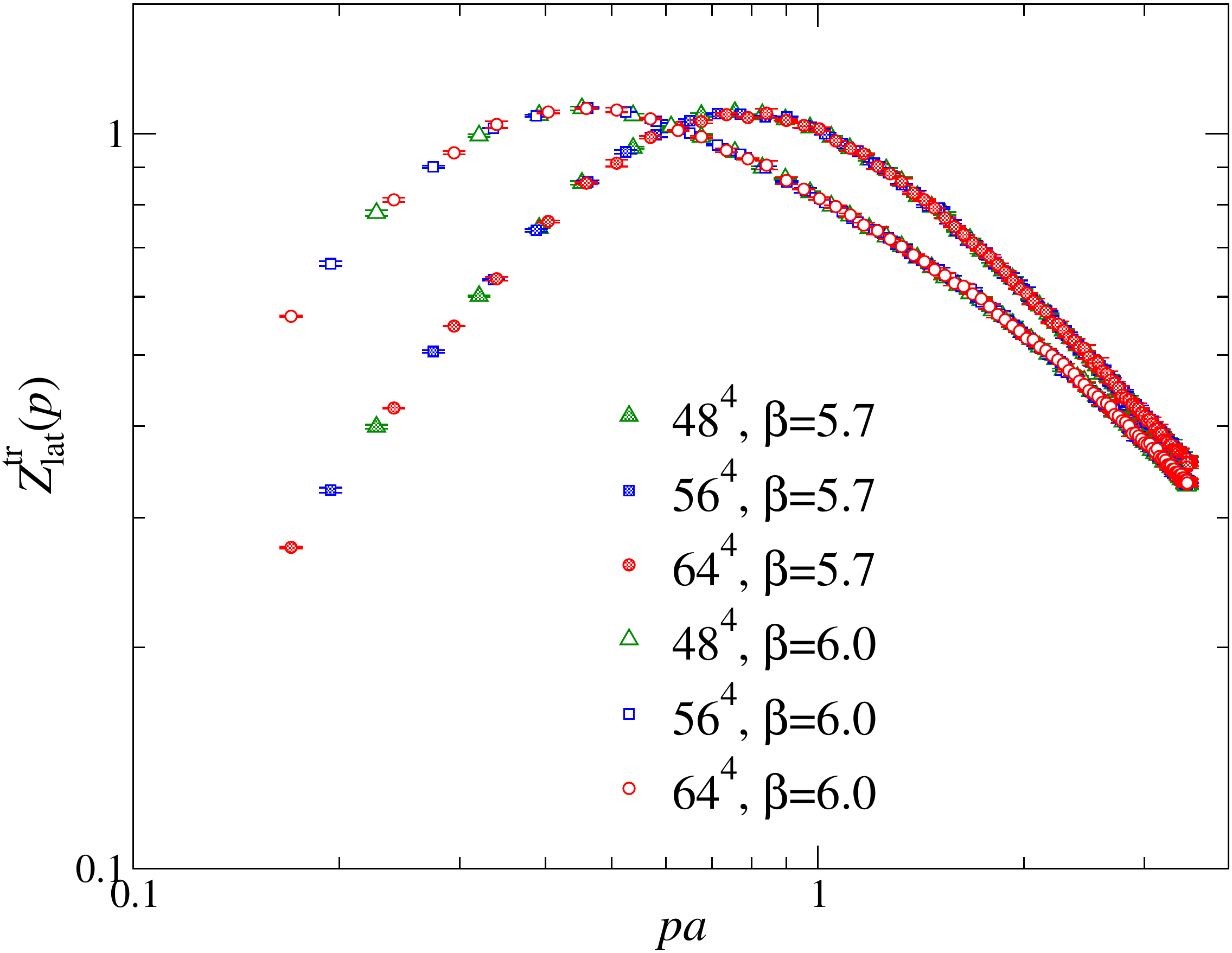}}
\end{center}\end{minipage}
\hspace{0.03\hsize}
\begin{minipage}{0.45\hsize}\begin{center}
\resizebox{1.\textwidth}{!}
{\includegraphics{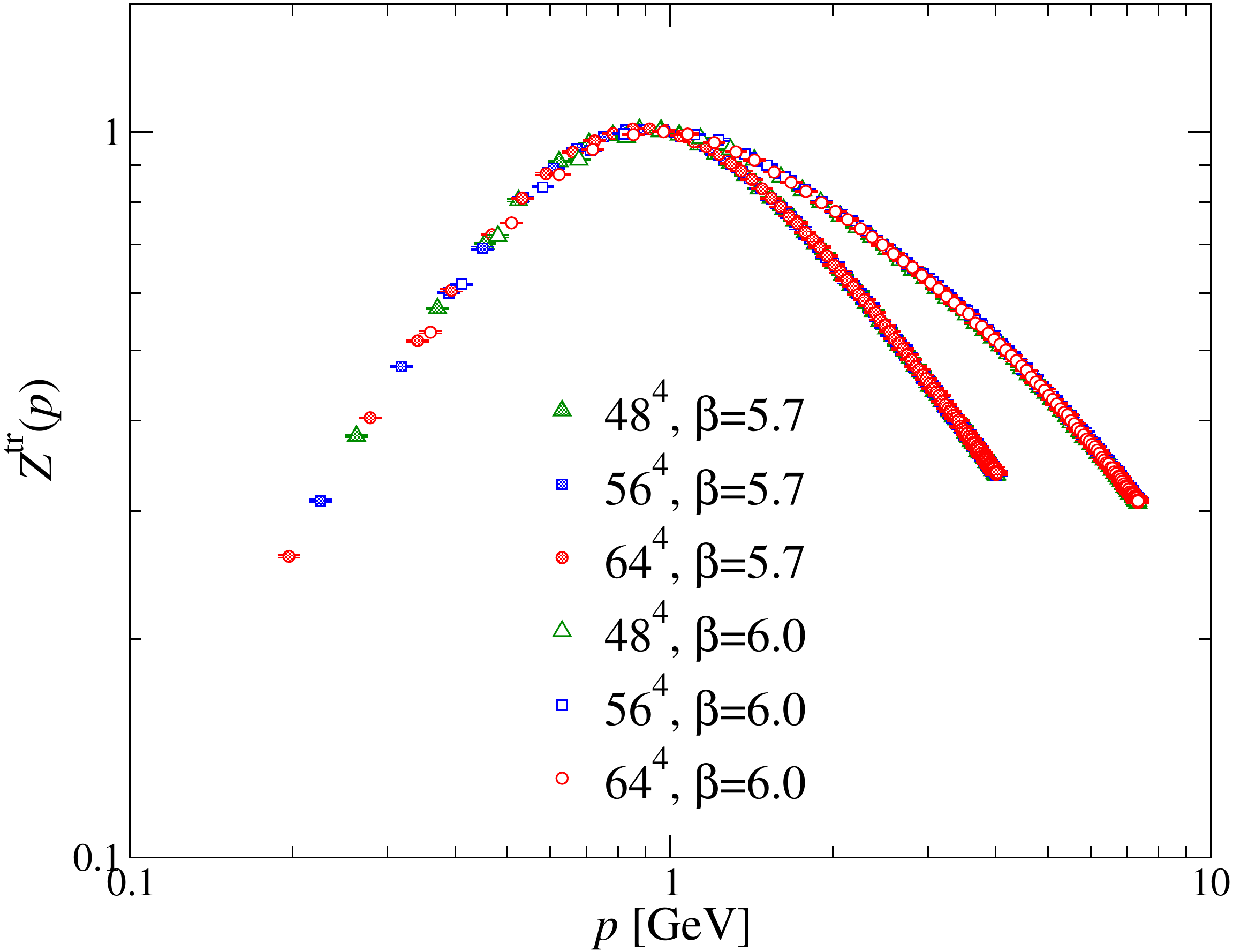}}
\end{center}\end{minipage}
\caption{
The dressing function of the unrenormalized instantaneous transverse gluon propagator,
$Z^{\rm tr}_{\rm lat}(\vec{p}) = \abs{\vec{p}}D^{\rm tr}_{\rm lat}(\vec{p})$,
at different lattice couplings in lattice units (left).
The dressing function renormalized at $\abs{\vec{p}}$=1 [GeV]
in physical units (right).
}
\label{fig:Ztr_unrnmlzd}
\end{figure*}

The left panel of \Fig{fig:Ztr_unrnmlzd} shows
the dressing function of the unrenormalized transverse gluon propagator
in lattice units.
It is apparent that the two curves cannot coincide by adopting
any scaling relation or by imposing any renormalization condition
since such manipulations correspond to the horizontal and the vertical
shifts in the log-log plot of the dressing function
but not the rotation or deformation of the curves.
Therefore, the scaling violation of the transverse gluon propagator
is purely due to discretization errors.


Furthermore, we observe that the IR behavior of the dressing function
at different couplings shows the same behavior, while
in the UV region the slope of the curves differs.
The right panel of \Fig{fig:Ztr_unrnmlzd}, in which the dressing function
renormalized at $\abs{p}$=1 [GeV] is plotted as a function of the physical momentum,
illuminates such a tendency;
the two curves almost fall on top of each other in the IR region
while the deviation between them is pronounced in the UV region.
This indicates that the scaling problem of $D^{\rm tr}$ resides
in the lattice data in the UV region.

This kind of behavior can be found for the instantaneous
free propagator; namely, the instantaneous propagator shows
the scaling violation even at the tree level (see \App{sec:Dtr_free}).
It is the crucial point of the scaling violation that
the instantaneous propagator is defined as the energy integral
of the unequal-time propagator.
For a finite temporal lattice spacing, the energy integral
is limited in the interval $[-\pi/a_{\tau},\pi/a_{\tau}]$,
and it induces a spurious $\abs{\vec{p}}$ dependence
on the instantaneous propagator.
It leads the discretization errors especially at large momenta,
as demonstrated in \App{sec:Dtr_free},
while it disappears only in the Hamiltonian limit $a_{\tau} \to 0$.
Accordingly, the lattice data in the UV region for
the instantaneous transverse gluon propagator suffer from
the discretization errors that cannot be eliminated with the cylinder cut
or the cone cut we applied.


One way to circumvent this problem is to exclude
the high momentum data from the analysis of the propagator.
This has been studied in 
\Ref{Nakagawa:2009zf}
combined with the matching procedure proposed in
\Ref{Leinweber:1998uu}
in order to find a reasonable value for the available momentum range.
It has been shown that the instantaneous transverse gluon propagator
shows scaling behavior by restricting the available momentum range.
The data on large lattices shown in \Fig{fig:Ztr_unrnmlzd}
illustrate that the discretization errors are relatively small
in the IR region, and this supports the validity of
the prescription in
\Ref{Nakagawa:2009zf}.
The another way is to calculate the unequal-time propagator
$D(\vec{p},p_4)$ and eliminates the $p_4$ dependence of
$D(\vec{p},p_4)$ as was discussed in
\Ref{Burgio:2008jr}, and the instantaneous transverse
gluon propagator has been shown to be multiplicatively renormalizable.

\section{Instantaneous temporal gluon propagator on the isotropic lattice}
\label{sec:D44_isotropic}


\begin{figure*}[htbp]
\begin{minipage}{0.45\hsize}\begin{center}
\resizebox{1.\textwidth}{!}
{\includegraphics{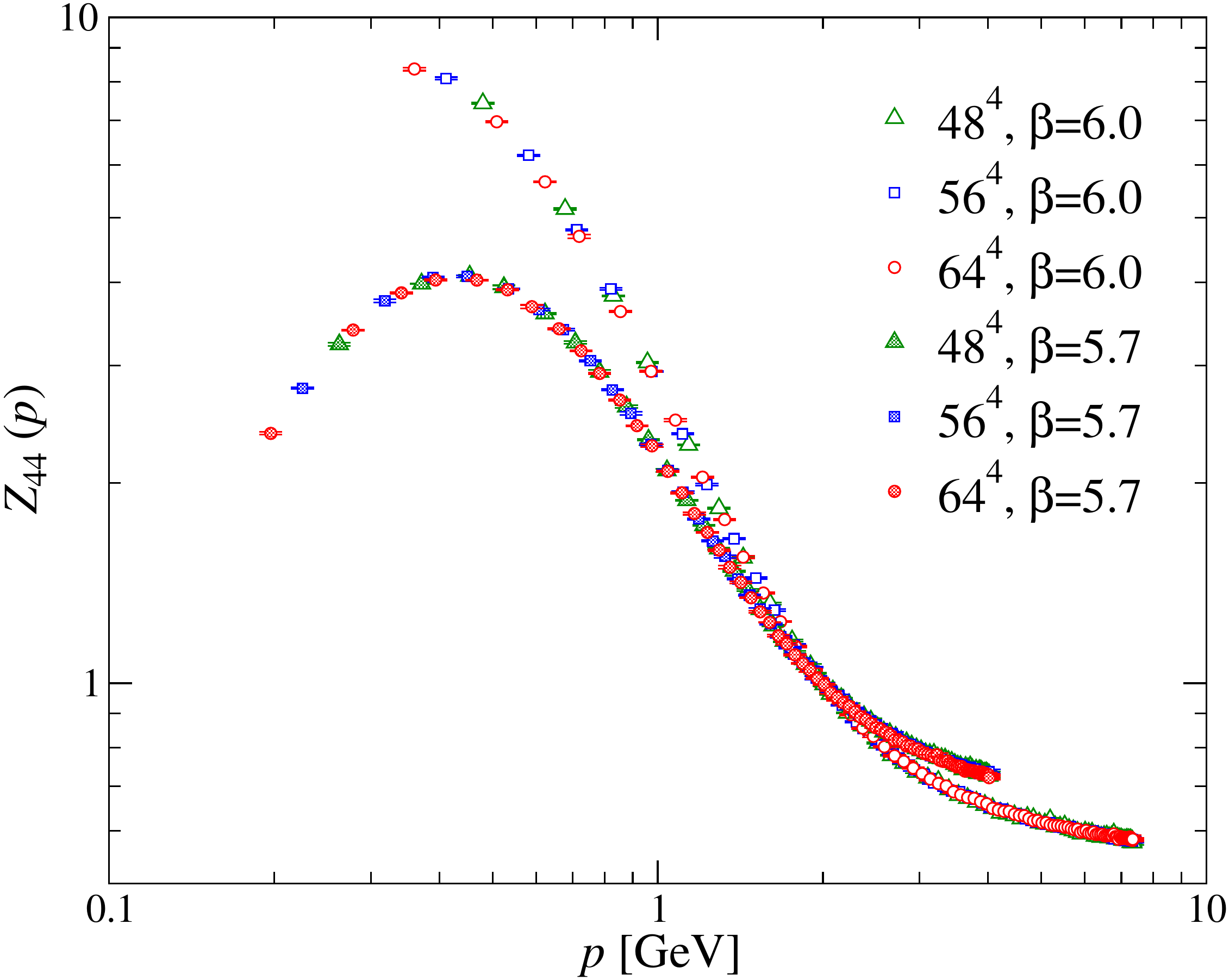}}
\end{center}\end{minipage}
\hspace{0.03\hsize}
\begin{minipage}{0.45\hsize}\begin{center}
\resizebox{1.\textwidth}{!}
{\includegraphics{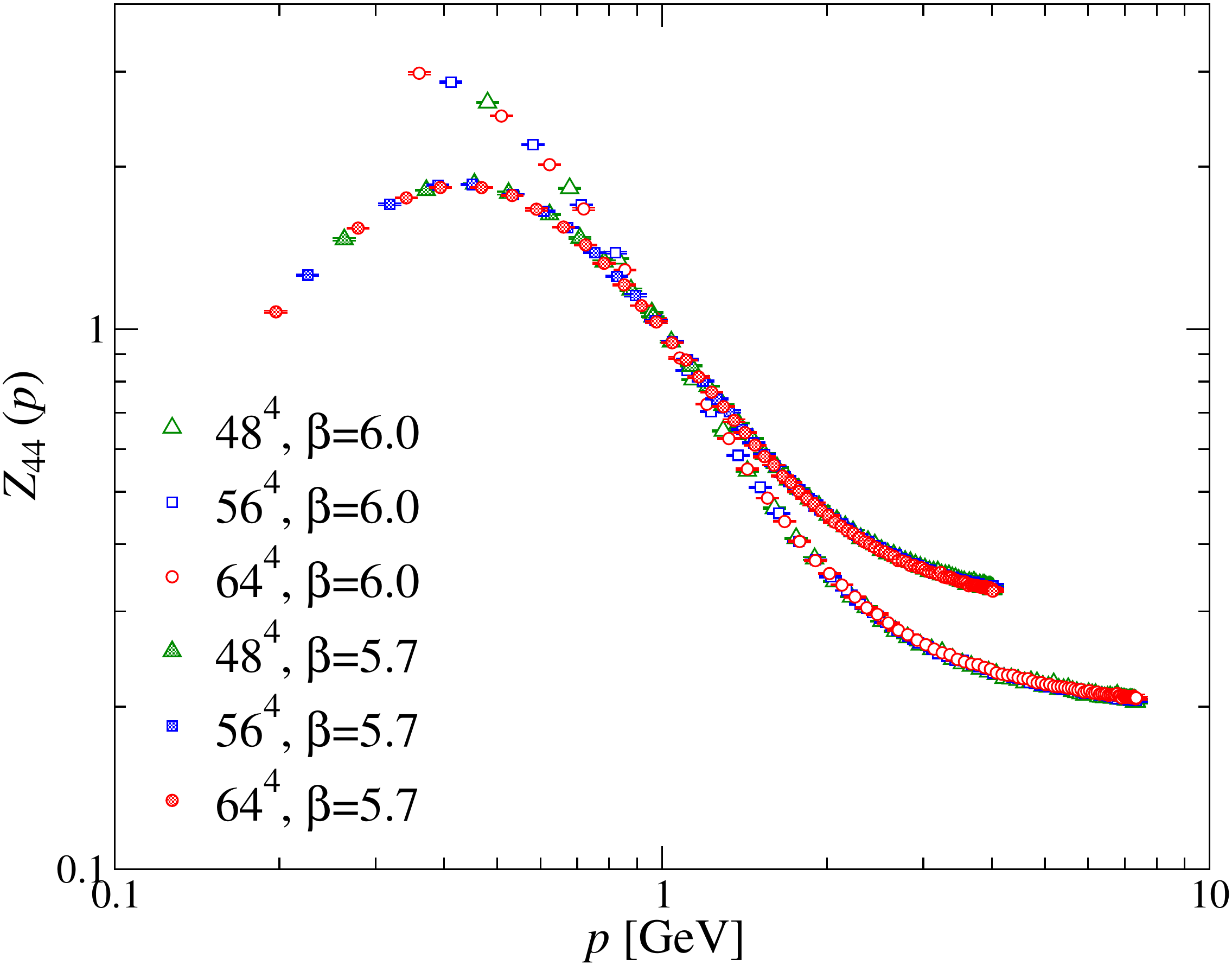}}
\end{center}\end{minipage}
\caption{
The dressing function of the instantaneous temporal gluon propagator
in physical units at $\beta=5.7$ and $\beta=6.0$.
The renormalization point is set to be 2 [GeV] in the left panel
and 1[GeV] in the right panel.
}
\label{fig:Z44_large_volume}
\end{figure*}

In this section, we discuss the isotropic lattice result
for the temporal gluon propagator.
The dressing function of the instantaneous temporal gluon propagator
on the isotropic lattice at $\beta$=5.7 and 6.0 is drawn
in \Fig{fig:Z44_large_volume}.
The dressing function is normalized such that
$Z_{44}(\abs{\vec{p}}=2{\rm [GeV]}) = 1$ in the left panel
and $Z_{44}(\abs{\vec{p}}=1{\rm [GeV]}) = 1$ in the right panel.
The dressing function $Z_{44}$ is constant for $\abs{\vec{p}}$
at the tree level.
In the Gribov-Zwanziger scenario, this is expected to diverge in
the IR limit resulting in the confining behavior of
the color-Coulomb potential, which is necessary condition
for color confinement in Coulomb gauge QCD.


Although the scaling violation is clearly visible
in $Z_{44}$ as in the transverse gluon propagator,
the scaling issue is worse for the temporal gluon propagator
than the transverse one.
In the case of $D^{\rm tr}$, the IR data do not much suffer
from the discretization errors and they almost fall on
top of each other by setting the renormalization point to
a small momentum.
Contrastingly, the discrepancy between two data sets of $Z_{44}$
corresponding to different lattice spacings remains
both in the IR and the UV regions
by changing the renormalization point
(compare the left and the right panels of \Fig{fig:Z44_large_volume}).
Indeed, the scaling violation of $D_{44}$ is not settled by
the $\alpha$-cut method employed in
\Ref{Nakagawa:2009zf},
indicating that the data at small momenta
may also be affected by the discretization errors.


Besides the scaling violation, we observe that the dressing function
is suppressed in the IR region.
For the temporal gluon propagator, we expect that its instantaneous
part behaves as $1/\abs{\vec{p}}^4$ in the IR region
since it corresponds to the color-Coulomb potential, which rises linearly
with distance between a quark and an antiquark
\cite{Greensite:2003xf,Nakamura:2005ux}.
As opposed to our expectation, the numerical results on the isotropic
lattice show that the dressing function $Z_{44}$ decreases
with momentum in the IR region for the coarser lattice.

The unexpected IR suppression of $Z_{44}$ may stem from
an incomplete isolation of the instantaneous part in $D_{44}$.
The unequal-time temporal gluon propagator can be decomposed
into the instantaneous part and the vacuum polarization part as
\Eq{eq:uneqD44} in the continuum theory, and it gives
\begin{equation}
\int^{\epsilon}_{-\epsilon} dt D_{44}(\vec{x},t) = V(\vec{x}) + O(\epsilon).
\end{equation}
On a lattice with finite $a_{\tau}$, $\epsilon$ is of the order of $a_{\tau}$
and we would have a $O(a_{\tau})$ contribution from the polarization term
in the instantaneous $D_{44}$ on the lattice.
The unwanted suppression at small momenta may originate from
such a contribution of the polarization term, and furthermore,
the energy integral of the polarization term
would produce a spurious $\abs{\vec{p}}$ dependence
as for the transverse gluon propagator, which leads the scaling violation.

Before closing the section, we summarize the points so far:
(1) The transverse gluon propagator on the isotropic lattice
suffers from the discretization errors at large momenta and
it can be ascribed to the spurious $\abs{\vec{p}}$ dependence
coming from the temporal lattice cutoff in the energy integral
defining the instantaneous propagator.
(2) The temporal gluon propagator is affected both in the IR
and the UV regions by the discretization errors, and
it may stem from the $O(a_{\tau})$ contribution of the polarization term
in the instantaneous propagator besides the limited energy integral.

\section{Instantaneous transverse gluon propagator on anisotropic lattices}
\label{sec:Dtr_anisotropic}


\begin{figure*}[htbp]
\begin{minipage}{0.45\hsize}\begin{center}
\resizebox{1.\textwidth}{!}
{\includegraphics{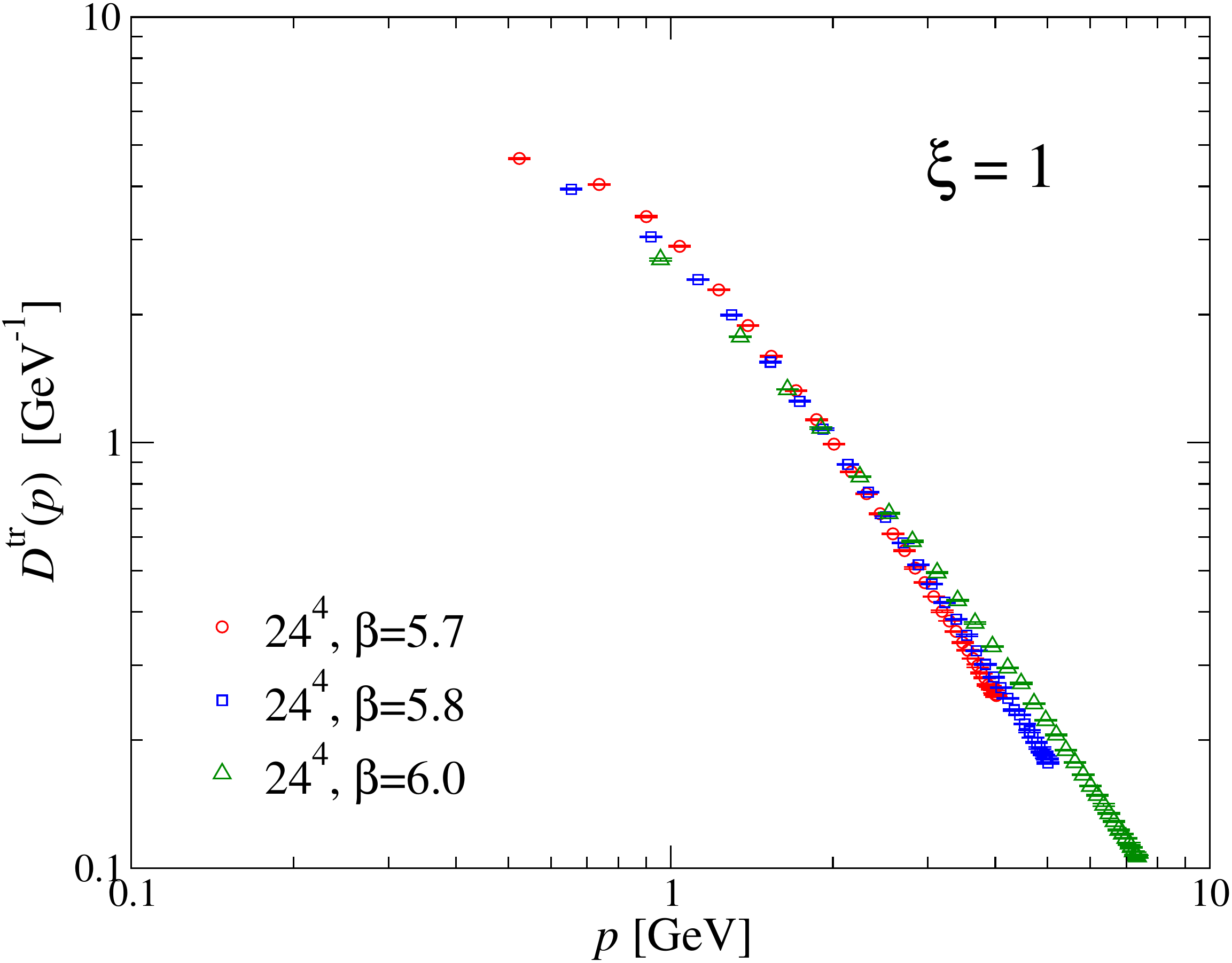}}
\resizebox{1.\textwidth}{!}
{\includegraphics{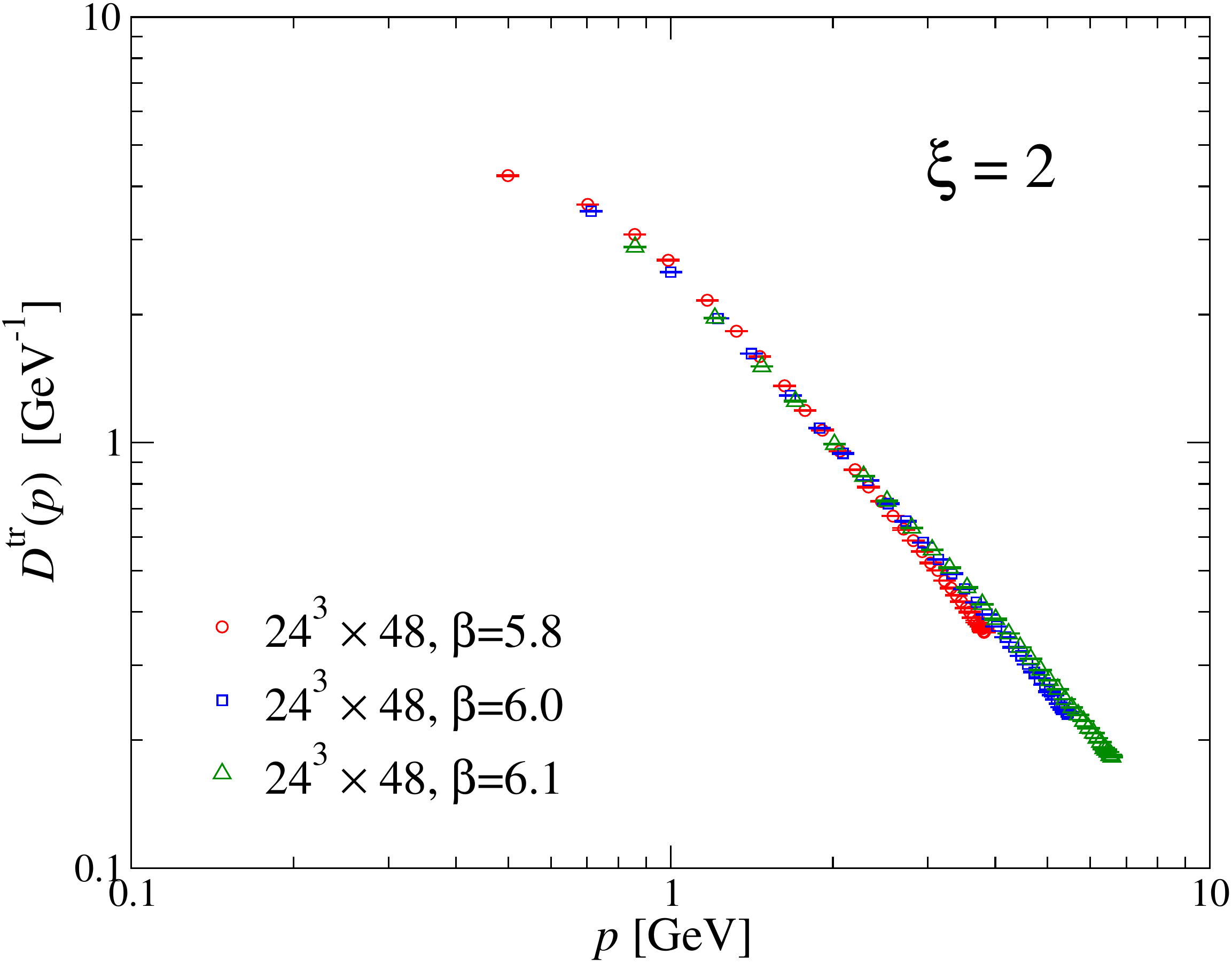}}
\end{center}\end{minipage}
\hspace{0.03\hsize}
\begin{minipage}{0.45\hsize}\begin{center}
\resizebox{1.\textwidth}{!}
{\includegraphics{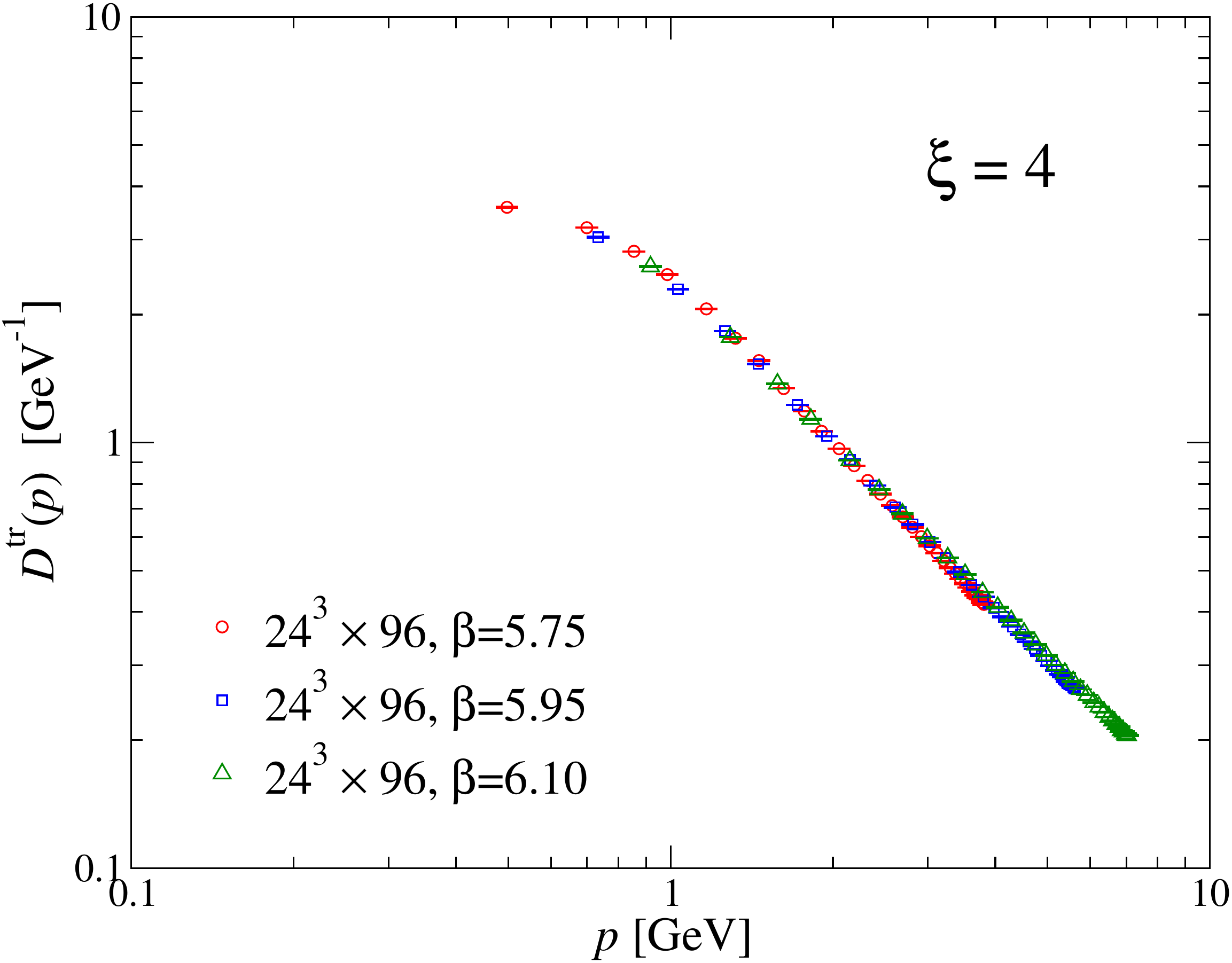}}
\resizebox{1.\textwidth}{!}
{\includegraphics{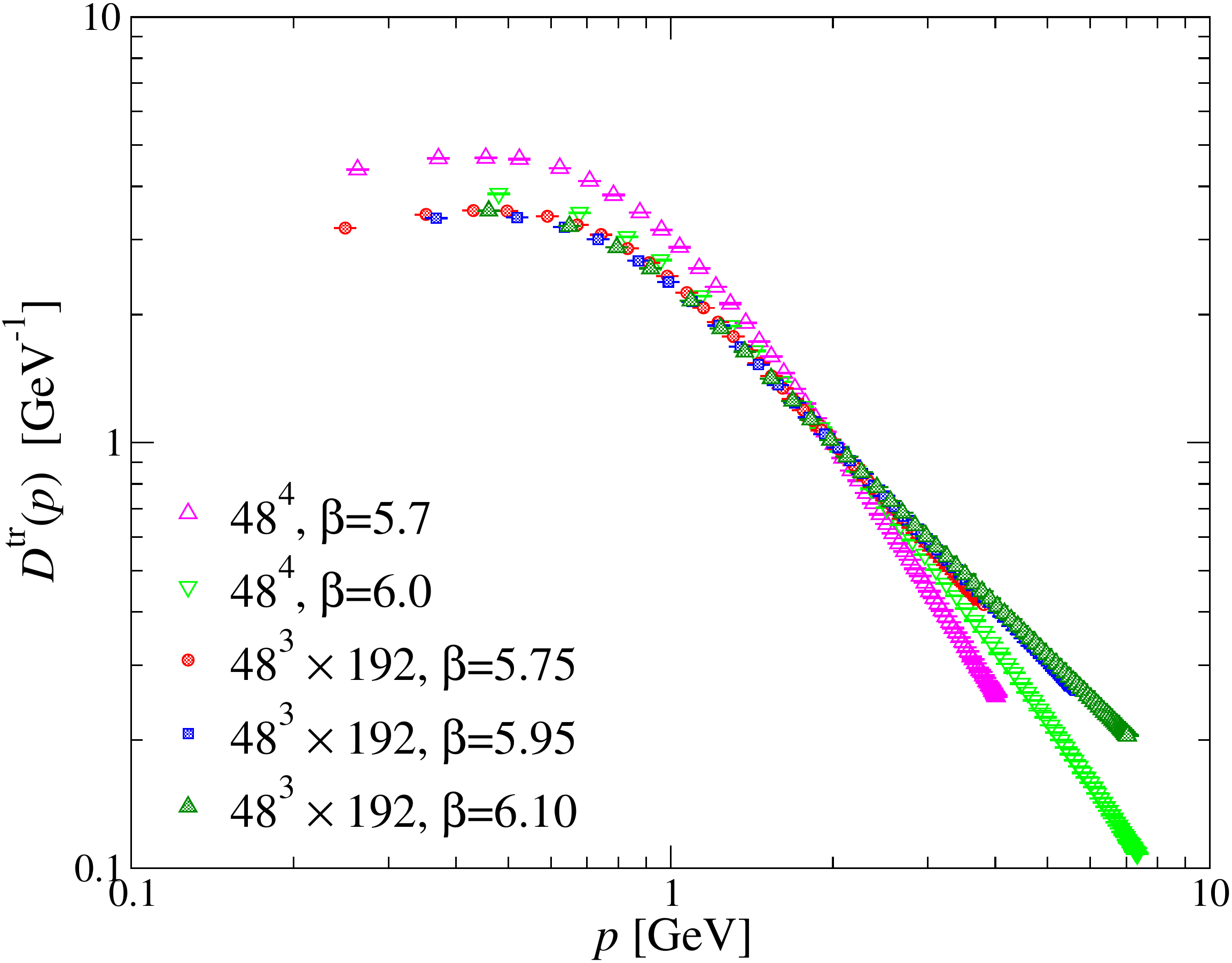}}
\end{center}\end{minipage}
\caption{
The instantaneous transverse gluon propagator
on the isotropic lattice (top left),
on the anisotropic lattices with $\xi=2$ (bottom left),
and with $\xi=4$ (top right).
The results for the isotropic lattice and the anisotropic lattice
with $\xi=4$ on the large lattice volume are drawn together
in one figure for direct comparison
(bottom right).
The propagator is renormalized to unity at $\abs{\vec{p}}=2$ [GeV].
}
\label{fig:Dtr_renormalized}
\end{figure*}

The instantaneous transverse gluon propagators for several anisotropies
are drawn in \Fig{fig:Dtr_renormalized}.
Compared to the isotropic case,
the deviations of three curves corresponding to different lattice couplings
become moderate on the anisotropic lattice with $\xi=2$.
Further increase of $\xi$ leads to a nice scaling behavior
and the data points for $\xi=4$ almost fall on top of one curve.
Accordingly, our results on the anisotropic lattices support
our expectation that scaling violation observed in the instantaneous
transverse gluon propagator disappears in the limit $\xi\to\infty$
($a_{\tau} \to 0$).


In order to investigate how scaling violation is cured
by getting close to the Hamiltonian limit, we quantify
the difference of two curves by the following function,
\begin{equation}\label{chi2}
\chi^2 = \sum_{i=1}^{n_{\rm f}}
\left( \frac{D_{\rm f}(p_i)-D_{\rm c}^{\rm int}(p_i)}{\sigma_{{\rm f},i}} \right)^2
+ \sum_{i=1}^{n_{\rm c}}
\left( \frac{D_{\rm c}(p_i)-D_{\rm f}^{\rm int}(p_i)}{\sigma_{{\rm c},i}} \right)^2.
\end{equation}
$D_{\rm f}(p_i)$ represents the measured value of the propagator at
the momentum $p_i$ and $\sigma_{{\rm f},i}$ the corresponding statistical error.
$D_{\rm c}^{\rm int}(p_i)$ is the estimated value
obtained by a cubic spline interpolation of the other lattice data set $D_{\rm c}$.
The subscripts f and c label two different data sets, meaning ``finer''
and ``coarser''.
The summations extend over the data points at which the two lattice data sets
overlap in the momentum.

We compute $\chi^2$ defined above for the following data sets:
\begin{center}
\begin{tabular}{clcl}
(1) \quad & $(24^3 \times 48, \beta=5.80, \xi=2)$ &
and & $(24^3 \times 48, \beta=6.00, \xi=2)$ \\
(2) \quad & $(24^3 \times 96, \beta=5.75, \xi=4)$ &
and & $(24^3 \times 96, \beta=5.95, \xi=4)$.
\end{tabular}
\end{center}
We note that the physical volumes are very similar;
(1) $V \sim 4.29^4$ [fm$^4$] and $V \sim 2.94^4$ [fm$^4$], and
(2) $V \sim 4.31^4$ [fm$^4$] and $V \sim 2.92^4$ [fm$^4$], respectively.
For each case, we found
\begin{center}
\begin{tabular}{ll}
(1) \quad & $\chi^2/N_{\rm DF} = 867$ \\
(2) \quad & $\chi^2/N_{\rm DF} = 309$,
\end{tabular}
\end{center}
where $N_{\rm DF}$ is the number of degrees of freedom
of the $\chi^2$ analysis.
Decreasing the temporal lattice spacing by a factor of 2,
$\chi^2/N_{\rm DF}$ reduces by about a factor 3.
We note that the absolute value of $\chi^2$ is unimportant
since it depends on the absolute value of the propagator,
which can take an arbitrarily large (or small) value
by multiplicative renormalization.
$\chi^2$ defined above make sense only when we compare
the $\chi^2$ values under the same renormalization condition
and the fixed ratio of the physical volumes of the lattice data sets
to be analyzed.


In the right bottom panel of \Fig{fig:Dtr_renormalized},
the instantaneous transverse gluon propagator
on the spatial lattice extent $L_{\sigma}=48$ is plotted
both for the isotropic lattice and the $\xi=4$ anisotropic lattice.
We observe that the propagator has a maximum at about 500 [MeV]
irrespective of the lattice coupling and the anisotropy,
and it decreases with the momentum in the IR region.
Accordingly, the turnover of the transverse gluon propagator
survives in the Hamiltonian limit.


\begin{figure}[htbp]
\begin{center}
\resizebox{0.45\textwidth}{!}
{\includegraphics{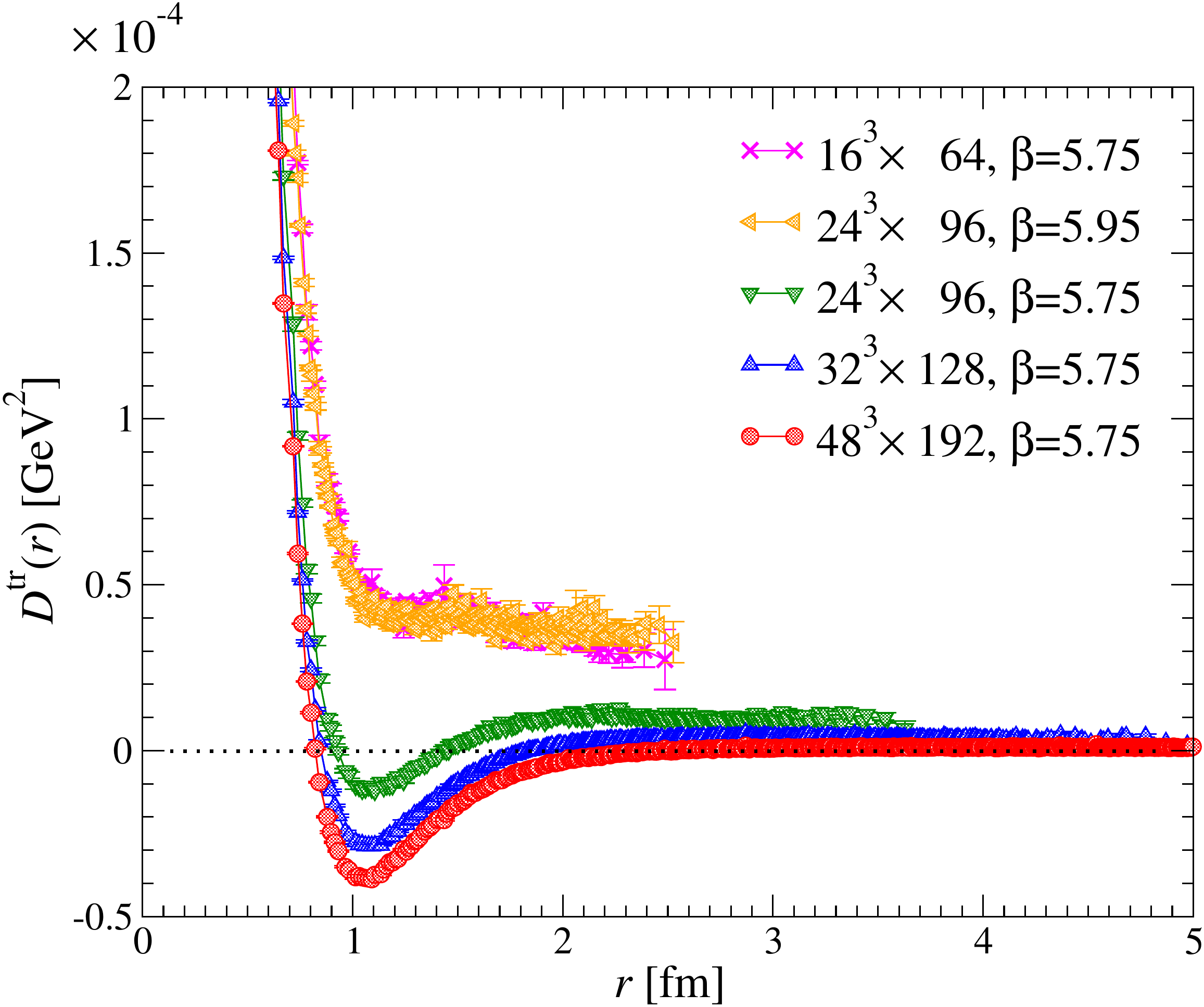}}
\end{center}
\caption{
The unrenormalized instantaneous gluon propagator in position space
(the correlation function of the gauge fields)
on $\xi=4$ anisotropic lattice.
The physical volumes are
$2.87^4$ [fm$^4$] (crosses),
$2.92^4$ [fm$^4$] (left triangles),
$4.31^4$ [fm$^4$] (down triangles),
$5.74^4$ [fm$^4$] (up triangles), and
$8.61^4$ [fm$^4$] (circles).
}
\label{fig:Dtr_xspace}
\end{figure}

The confined behavior of the gluon propagator can be seen
more directly in position space.
The unrenormalized correlation function of the transverse gauge fields
on the anisotropic lattice $\xi=4$ is depicted in \Fig{fig:Dtr_xspace}.
On small lattices (crosses and left triangles in the figure),
the correlation function is positive and a decreasing function
of the distance.
As the physical volume increases, $D^{\rm tr}(r)$ develops
a dip at about $r=1$ [fm] and becomes negative around the dip.
On the largest volume, the correlation function quickly decreases
with distance at small $r$ and becomes negative in the range
$r \sim 1-2$ [fm] and then vanishes at large distances.
It means that the gauge fields have no correlation over the hadronic scale
in sharp contrast to the massless particles for which the correlation
function behaves as $1/r^2$.
We notice that the results at the approximately fixed physical volume
nicely agree with each other; namely, lattice discretization errors
do not seriously affect the correlation function on the anisotropic lattice.

\section{Power law fitting of $D^{\rm tr}$ at small and large momenta}
\label{sec:Dtr_anisotropic_fit}


We note that the momentum dependence of the instantaneous transverse
gluon propagator on the anisotropic lattices in the UV region
significantly differs from that on the isotropic lattice
(see the bottom right panel in \Fig{fig:Dtr_renormalized}).
On the isotropic lattice, the slope at high momenta
gets smaller as the lattice couplings increase.
On the anisotropic lattices, the slope decreases further.
In order to investigate the UV behavior of the transverse
gluon propagator, we make the power law ansatz,
\begin{equation}
D^{\rm tr}(\vec{p}) = \frac{d_1}{\abs{\vec{p}}^{1+\gamma_{\rm gl}^{\rm UV}}},
\end{equation}
and fit this ansatz to the data on $24^3 \times L_{\tau}$ lattices
in the momentum range $\abs{\vec{p}} \ge 6$ [GeV].
The fitted parameters are given in \Tab{tab:fit_Dtr_UV}.
We see that $\gamma_{\rm gl}^{\rm UV}$ gets small values
as the anisotropy increases, and that for $\xi=4$
takes about one third of that on the isotropic lattice.

As we have discussed in \Sec{sec:Dtr_isotropic},
the inverse of the instantaneous propagator can be interpreted
as the energy dispersion relation.
Since gluons are expected to behave as free massless particles
at sufficiently large momentum due to asymptotic freedom,
this interpretation can be allowed if the instantaneous transverse
gluon propagator behaves as $1/\abs{\vec{p}}$ in the UV region,
which corresponds to the null UV exponent.
Therefore, the decrease of $\gamma_{\rm gl}^{\rm UV}$ with increasing $\xi$
is consistent with the asymptotic free field behavior
of the gluon fields in the continuum limit
\footnote{
At least, the linear extrapolation of $\gamma_{\rm gl}^{\rm UV}$ to the Hamiltonian
limit gives $\gamma_{\rm gl}^{\rm UV}(\xi\to\infty)=0.0852(201)$
and $\chi^2/N_{\rm DF}=11.1$, that is, the value of $\gamma_{\rm gl}^{\rm UV}$
decreases but remains finite in this analysis with three anisotropies,
although the linear extrapolation may be too naive and inadequate, and
a careful study of taking the Hamiltonian limit is needed.
}.

\begin{table*}[htdp]
\caption{
The result of the UV power law fitting of $D^{\rm tr}$
in the momentum range $\abs{\vec{p}} \ge 6$ [GeV].
}
\begin{center}\begin{tabular}{cccc}
\hline\hline
$( L_{\sigma}, L_{\tau}, \xi, \beta )$ & $d_1$ &
$\gamma_{\rm gl}^{\rm UV}$ & $\chi^2/N_{\rm DF}$ \\ 
\hline
(24, 24, 1, 6.0) & 4.89(14) & 0.920(15) & 1.34 \\
(24, 48, 2, 6.1) & 3.83(25) & 0.617(35) & 0.86 \\
(24, 96, 4, 6.1) & 2.49(7)  & 0.282(15) & 0.68 \\
\hline\hline
\end{tabular}\end{center}
\label{tab:fit_Dtr_UV}
\end{table*}


In order to explore the IR behavior of the instantaneous transverse
gluon propagator, we make the power law ansatz,
\begin{equation}
D^{\rm tr}(\vec{p}) = d_2 \abs{\vec{p}}^{\gamma_{\rm gl}^{\rm IR}},
\end{equation}
in the IR region.
The fitted parameters are listed in \Tab{tab:fit_Dtr_IR}.

Although the fitting becomes worse and the IR exponent
$\gamma^{\rm IR}_{\rm gl}$ becomes small as the maximum momentum
of the fitting range increases,
$\gamma^{\rm IR}_{\rm gl}$ takes a positive value in all the cases.
For the anisotropic lattice, we need much larger lattices to extract
the IR exponent with 	an acceptable $\chi^2$ value.
In both the isotropic and the anisotropic cases,
our result of the IR fitting predicts the vanishing transverse
gluon propagator at zero momentum.
Given the fact that the fitted values of $\gamma^{\rm IR}_{\rm gl}$
increases with decreasing the maximum momentum of the fitting range,
the $\gamma^{\rm IR}_{\rm gl}$ values listed in \Tab{tab:fit_Dtr_IR}
would give a lower bound for the IR exponent of the transverse
gluon propagator in the Coulomb gauge.

We note that fitting the transverse gluon propagator
with the Gribov-type ansatz, \Eq{eq:GribovAnsatz},
which successfully reproduced the lattice data in
\Ref{Burgio:2008jr},
did not work in our lattice data for both the isotropic
and the anisotropic cases.
At the same time, 
the peak position of the transverse gluon propagator
differs between our results and the results in
\Ref{Burgio:2008jr};
it is about 500 [MeV] in our case and
the fitting analysis in
\Ref{Burgio:2008jr}
gives 880(10) [MeV], which is rather close to the peak
position of the dressing function $Z^{\rm tr}$ in our results
(see the right panel of \Fig{fig:Ztr_unrnmlzd}).
It deserves further study to clarify whether this discrepancy
comes from the gauge group [SU(3) in this study and SU(2) in
\Ref{Burgio:2008jr}] or the adopted prescriptions
to circumvent scaling violation.

\begin{table*}[htdp]
\caption{
The result of the IR power law fitting of $D^{\rm tr}$.
$\abs{\vec{p}_{max}}$ represents
the maximum momentum of the fitting range.}
\label{tab:fit_Dtr_IR}
\begin{center}\begin{tabular}{ccccc}
\hline\hline
$(L_{\sigma}, \xi, \beta)$ &
$\abs{\vec{p}_{max}}$ & $d_2$ & $\gamma_{\rm gl}^{\rm IR}$ & $\chi^2/N_{\rm DF}$ \\
\hline
& 0.27 & 6.64(17) & 0.311(17) & 0.274 \\
( 48 $\sim$ 64, 1, 5.7 )
& 0.30 & 6.24(9) & 0.271(10) & 4.48 \\
& 0.33 & 5.98(6) & 0.241(8) & 7.82 \\
\hline
( 48, 4, 5.75 )
& 0.45 & 4.09(2) & 0.174(4) & 94.3 \\
\hline\hline
\end{tabular}\end{center}
\end{table*}

\section{Instantaneous temporal gluon propagator on anisotropic lattices}
\label{sec:Z44_anisotropic}


\begin{figure*}[htbp]
\begin{minipage}{0.45\hsize}\begin{center}
\resizebox{1.\textwidth}{!}
{\includegraphics{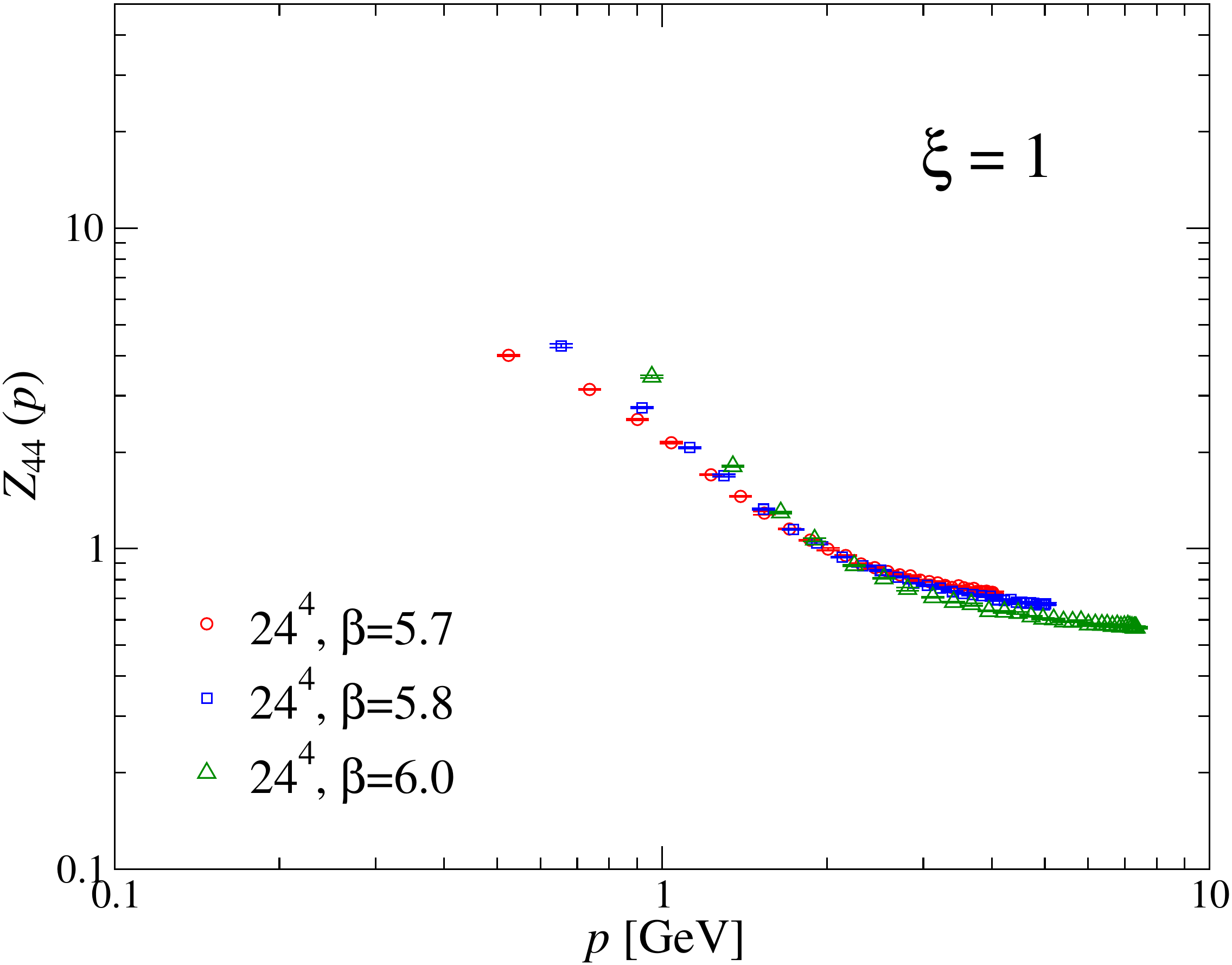}}
\resizebox{1.\textwidth}{!}
{\includegraphics{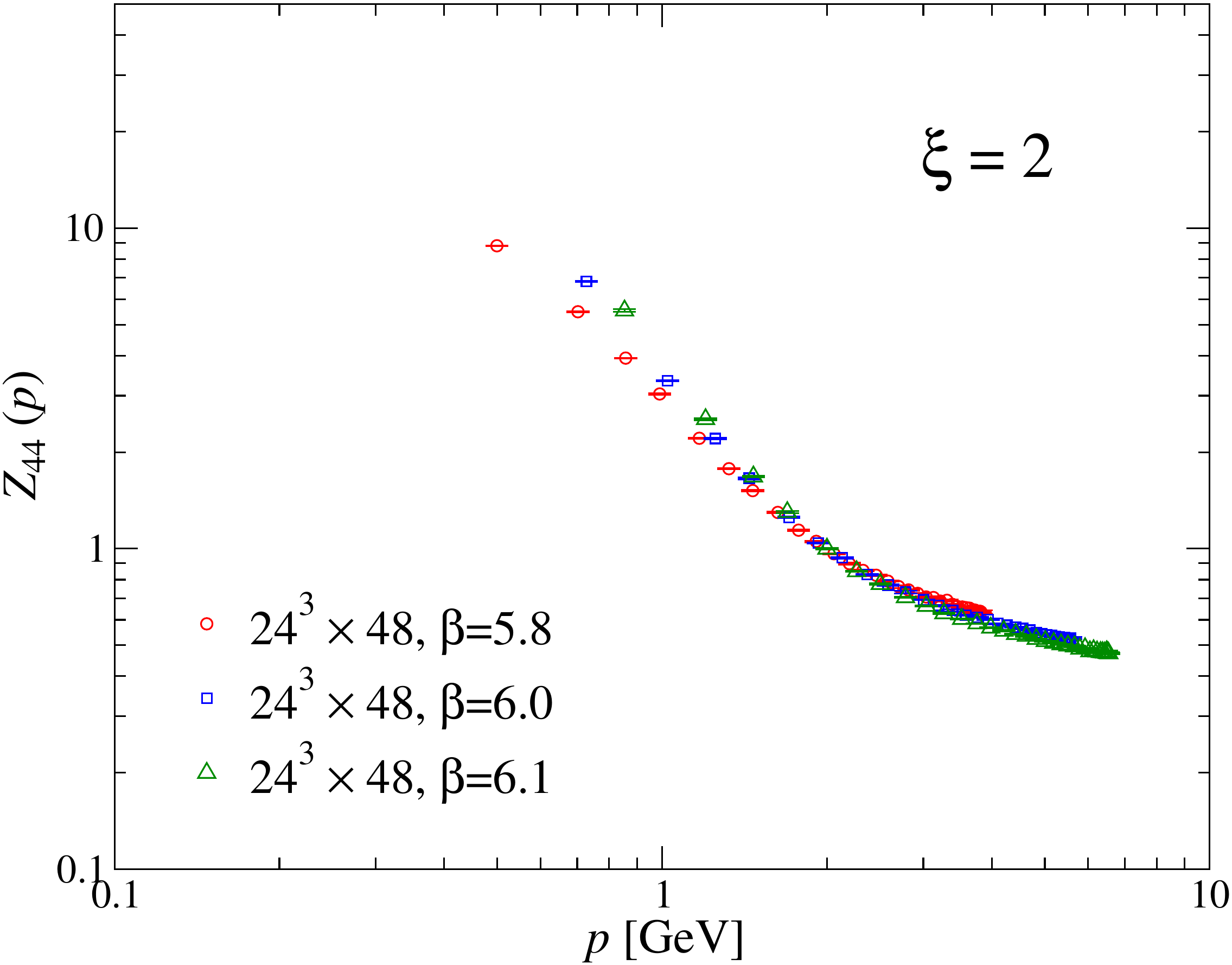}}
\end{center}\end{minipage}
\hspace{0.03\hsize}
\begin{minipage}{0.45\hsize}\begin{center}
\resizebox{1.\textwidth}{!}
{\includegraphics{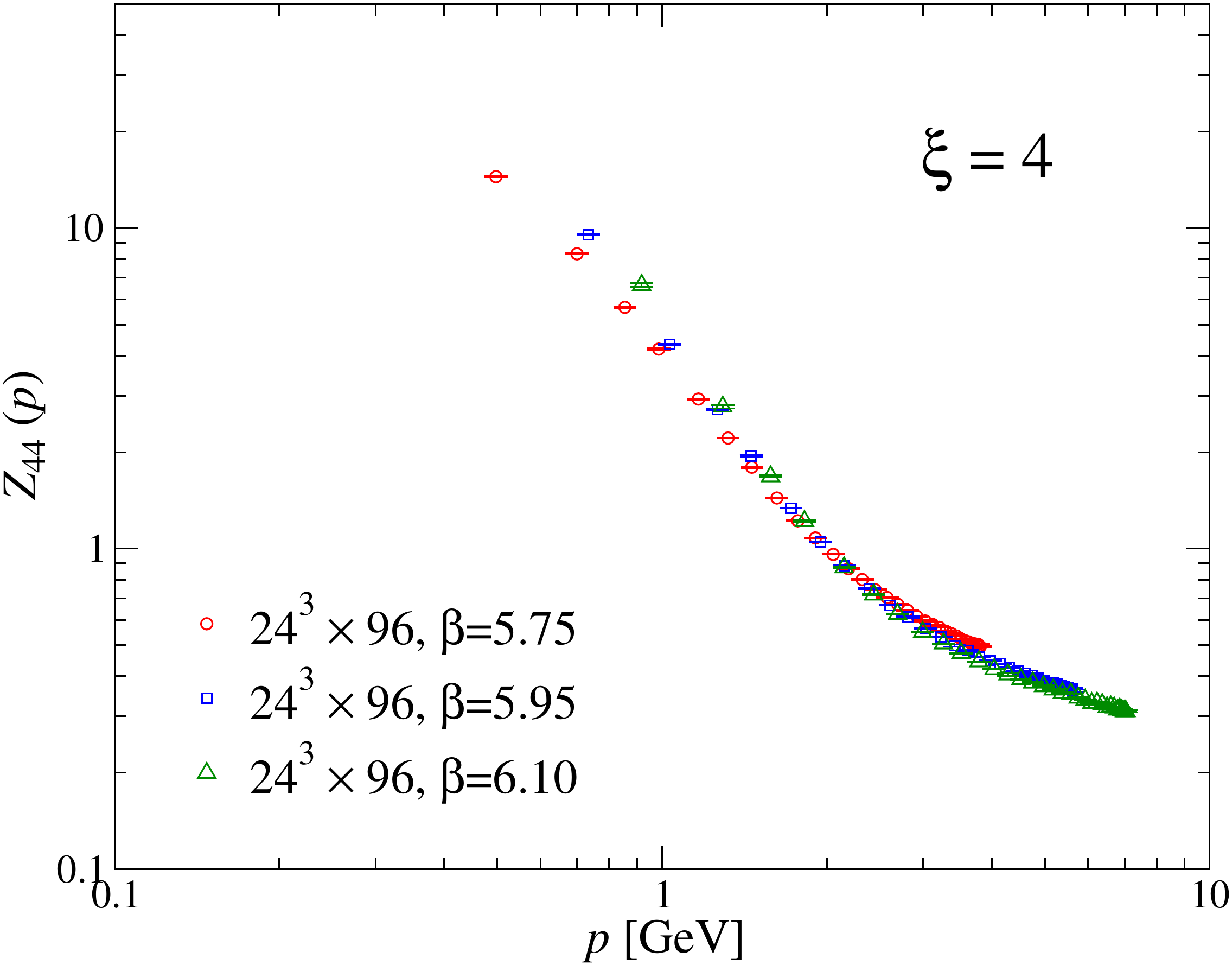}}
\resizebox{1.\textwidth}{!}
{\includegraphics{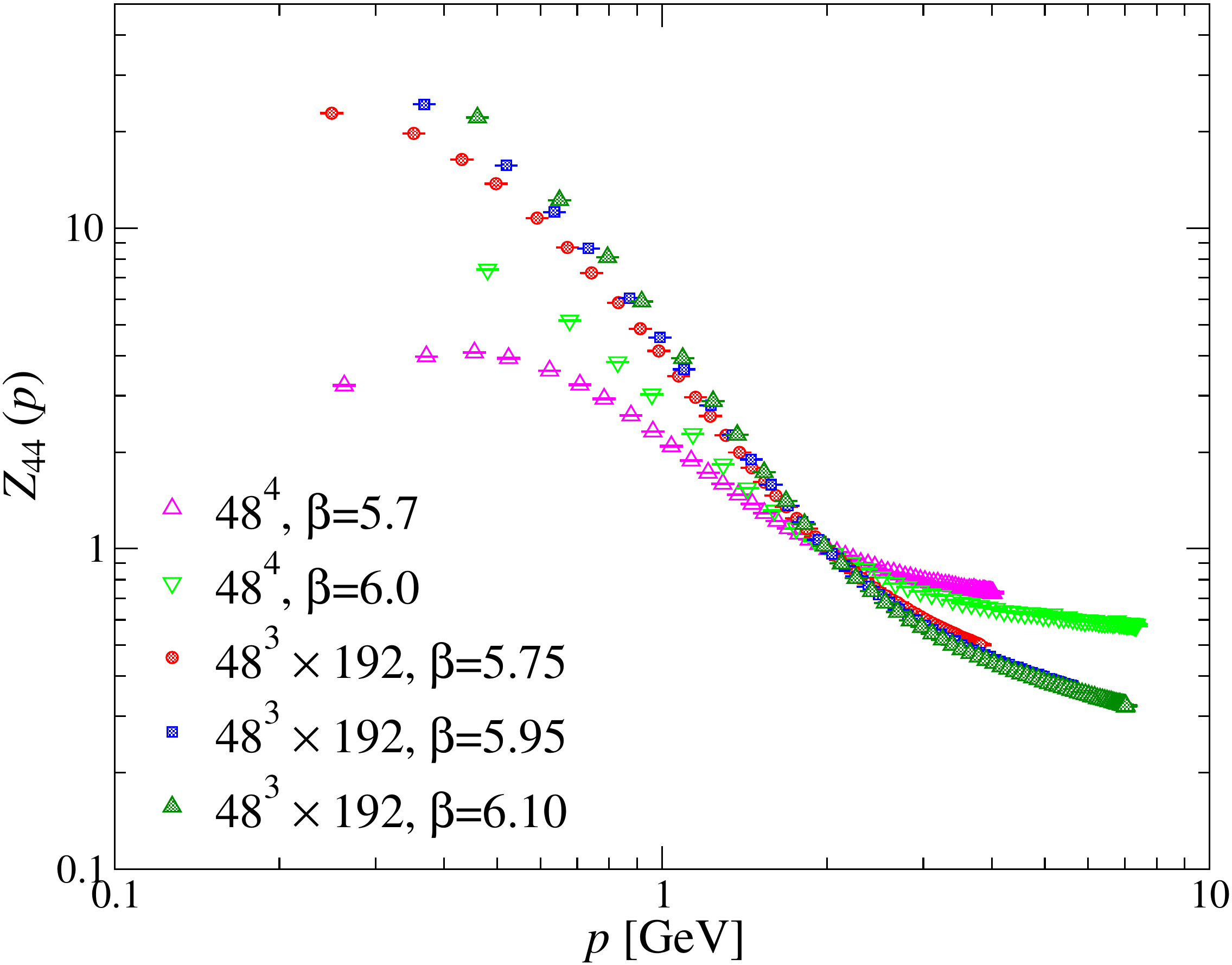}}
\end{center}\end{minipage}
\caption{
The dressing function of the instantaneous temporal gluon propagator
on the isotropic lattice (top left),
on the anisotropic lattice with $\xi=2$ (bottom left),
and that with $\xi=4$ (top right).
The results for the isotropic lattice and the anisotropic lattice
with $\xi=4$ on the large lattice volume are drawn together
in one figure for direct comparison
(bottom right).
The dressing function is renormalized to unity at $\abs{\vec{p}}=2$ [GeV].
}
\label{fig:Z44_renormalized}
\end{figure*}

We next discuss the instantaneous temporal gluon propagator,
which is related to the color-Coulomb potential.
The dressing function of
the time-time component of the gluon propagator is shown
in \Fig{fig:Z44_renormalized} for the isotropic lattice
and the anisotropic lattices.
On the anisotropic lattices, the dressing function shows
much better scaling behavior than that on the isotropic lattice.
Although the small deviation can be seen both in the IR and
UV region, one can expect that the scaling behavior
is completely recovered in the Hamiltonian limit.


Moreover, we find that the IR behavior of $Z_{44}$
on the anisotropic lattice is different from that
on the isotropic lattice (see bottom right panel).
For the isotropic case, we see that the dressing function
increases slowly with decreasing the momentum,
and bends down on coarse lattice.
By contrast, $Z_{44}$ continues to rise on the anisotropic lattice
even for the coarsest lattice data ($\beta=5.75$), and
$Z_{44}$ at available smallest momentum for the anisotropic case
is about 10 times larger than that for the isotropic case.
We note that the spatial lattice spacing for $(\xi,\beta)=(4,5.75)$ is
larger than that for $(\xi,\beta)=(1,5.70)$.
This implies that $Z_{44}$ is very sensitive to the discretization effects, and
taking the Hamiltonian limit is crucial to cure scaling violation
for the temporal gluon propagator and to explore the genuine
IR divergent behavior in the Coulomb gauge QCD.

\section{Power law fitting of $Z_{44}$ at small and large momenta}
\label{sec:Z44_anisotropic_fit}


The asymptotic form of the color-Coulomb potential,
the instantaneous part of the temporal gluon propagator, is given by
\begin{equation}
\abs{\vec{p}}^2 V_{44}(\abs{\vec{p}})
\sim \frac{x_0}{2b_0\ln\left( \abs{\vec{p}}/\Lambda_{\rm Coul} \right)},
\end{equation}
where $x_0=12/11$ and $b_0=11/16\pi^2$ for SU(3) pure Yang-Mills theory,
and $\Lambda_{\rm Coul}$ is a finite QCD mass scale
\cite{Cucchieri:2000hv}.
We here fit the data for $Z_{44}$ with
\begin{equation}
Z_{44}(\abs{\vec{p}}) = Z
\frac{x_0}{2b_0\ln\left( \abs{\vec{p}}/\Lambda_{\rm Coul} \right)}
\end{equation}
in the momentum range $\abs{\vec{p}} > 6$ [GeV].
The fitted parameters $Z$ and $\Lambda_{\rm Coul}$ are
given in \Tab{tab:fit_Z44_UV} with $\chi^2/N_{DF}$.
We find that the fitted parameter $\Lambda_{\rm Coul}$
takes an unacceptably small value as a QCD mass scale,
although the $\chi^2$ value is reasonable for the isotropic lattice.
$\Lambda_{\rm Coul}$ increases with the anisotropy and
the fitted value $\Lambda_{\rm Coul}=0.8345(85)$ [GeV] for $\xi=4$
is the order of $\Lambda_{\rm QCD}$.
However, the numerical data for the $\Lambda_{\rm QCD}$ parameter are still unstable
under the increase of the anisotropy and $\Lambda_{\rm QCD}$ could change
by further varying $\xi$.
It can be stated that the anisotropy should be greater than $4$
otherwise we have physically inadequate results,
although it is difficult to estimate the Hamiltonian limit of $\Lambda_{\rm QCD}$.

\begin{table*}[htdp]
\caption{
The result of the UV logarithmic law fitting of $Z_{44}$
in the momentum range $\abs{\vec{p}} \ge 6$ [GeV].
}
\begin{center}\begin{tabular}{cccc}
\hline\hline
$( L_{\sigma}^3 \times L_{\tau}, \xi, \beta )$ & $Z$ &
$\Lambda_{\rm Coul}$ [GeV] & $\chi^2/N_{\rm DF}$ \\ 
\hline
(24$^4$,           1, 6.0) & 0.648(97) & 0.000968(1282) & 0.406 \\
(24$^3 \times$ 48, 2, 6.1) & 0.310(80) & 0.0385(507) & 1.20 \\
(24$^3 \times$ 96, 4, 6.1) & 0.0842(41)  & 0.845(85) & 2.22 \\
\hline\hline
\end{tabular}\end{center}
\label{tab:fit_Z44_UV}
\end{table*}


In the IR region, the color-Coulomb potential is expected
to behave as $1/\abs{\vec{p}}^2$, which gives a linearly rising potential
in position space.
Indeed, the lattice simulations of the color-Coulomb potential
obtained from the correlator of the partial Polyakov line revealed
such a linearity of the color-Coulomb potential at large distances
\cite{Greensite:2003xf,Nakamura:2005ux}.
Therefore, we fit the dressing function $Z_{44}$ with the power law form,
\begin{equation}
Z_{44}(\abs{\vec{p}}) = \frac{z}{\abs{\vec{p}}^{\gamma_{44}}}.
\end{equation}
Assuming that the contribution of the vacuum polarization term to
the instantaneous temporal gluon propagator is negligible,
we expect $\gamma_{44}=2$ giving a linearly rising color-Coulomb potential.

The fitted results are given in \Tab{tab:fit_Z44_IR}.
We observe that $\chi^2$ is extraordinary large and the fitting does not work.
Even though the scaling violation becomes moderate as the anisotropy
increases, we still have discretization effects which are visible
in \Fig{fig:Z44_renormalized}; the curves associated with different
lattice couplings deviate from each other in the IR region and
the slope gets steeper as the lattice spacing decreases.
Therefore, we are still not close to the Hamiltonian limit
where the scaling behavior is observed, and we cannot extract
the color-Coulomb string tension from $Z_{44}(\abs{\vec{p}})$
and compare it with that obtained from the link-link correlator.

\begin{table*}[htdp]
\caption{
The result of the IR power law fitting of $Z_{44}$.
$\abs{\vec{p}_{max}}$ represents
the maximum momentum of the fitting range.}
\begin{center}\begin{tabular}{ccccc}
\hline\hline
$(L_{\sigma}, L_{\tau}, \xi, \beta)$ &
$\abs{\vec{p}_{max}}$ & $z$ & $\gamma_{44}$ & $\chi^2/N_{\rm DF}$ \\
\hline
( 48, 48, 1, 6.0 )
& 0.90 & 3.08(1) & 1.22(1) & 265 \\
\hline
( 48, 192, 4, 6.1 )
& 0.90 & 5.42(2) & 1.84(1) & 248 \\
\hline\hline
\end{tabular}\end{center}
\label{tab:fit_Z44_IR}
\end{table*}

\section{Instantaneous $D_{44}$ in position space and the color-Coulomb potential}
\label{sec:D_44_and_Vc}


\begin{figure*}[htbp]
\begin{minipage}{0.45\hsize}\begin{center}
\resizebox{1.\textwidth}{!}
{\includegraphics{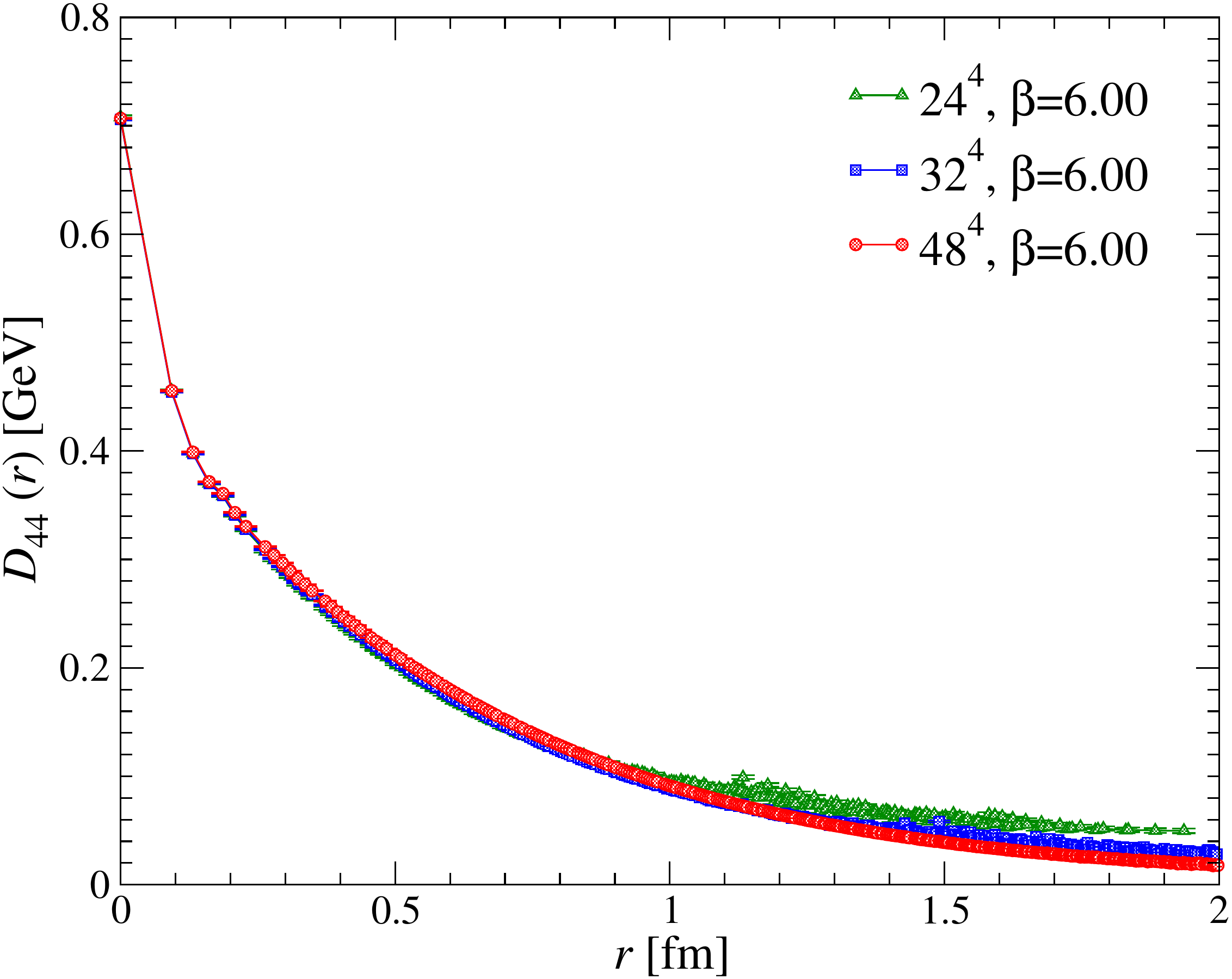}}
\end{center}\end{minipage}
\hspace{0.03\hsize}
\begin{minipage}{0.45\hsize}\begin{center}
\resizebox{1.\textwidth}{!}
{\includegraphics{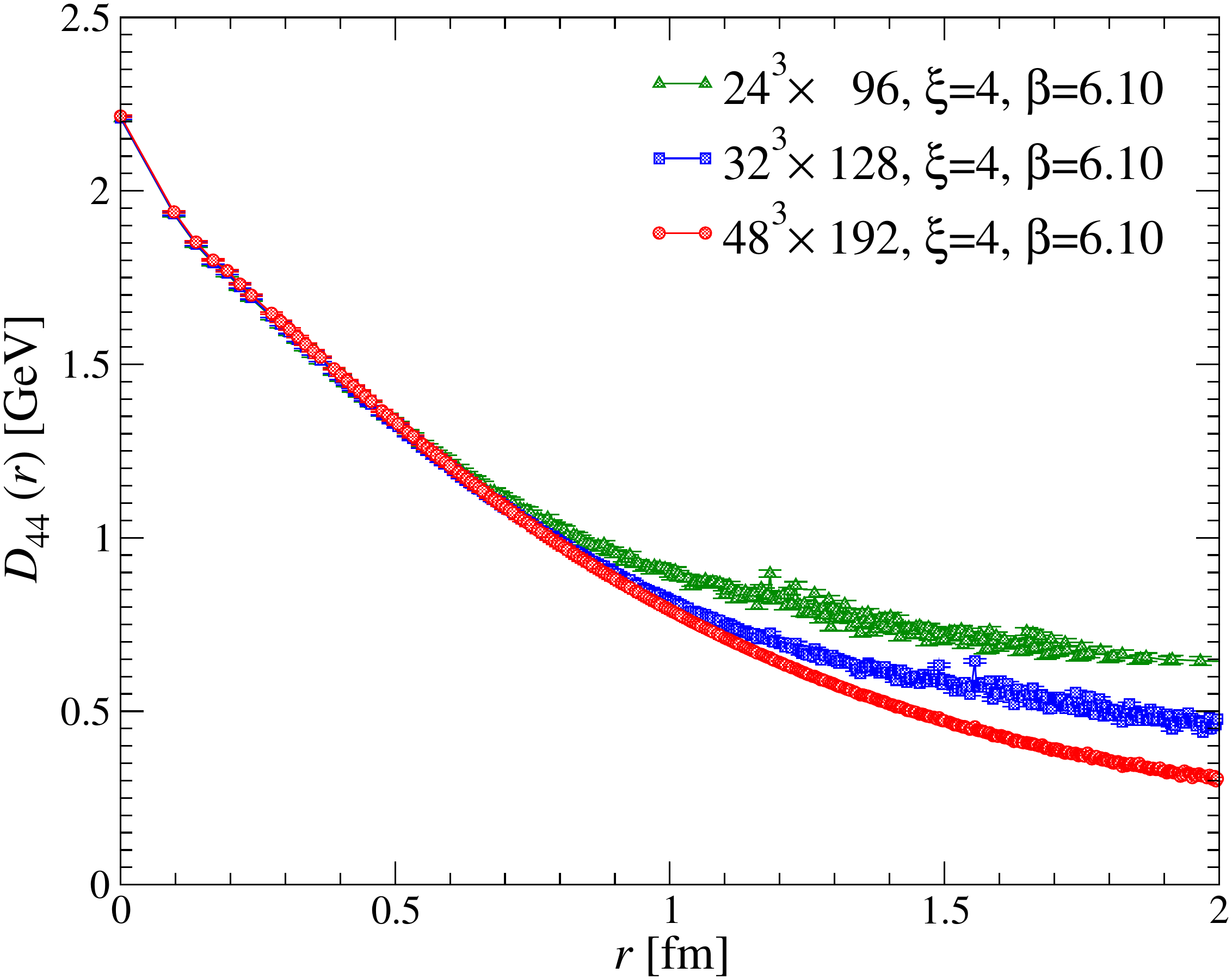}}
\end{center}\end{minipage}
\caption{
The instantaneous temporal gluon propagator in position space
on the isotropic lattice (left) and the $\xi=4$ anisotropic lattice (right).
}
\label{fig:D44_xspace}
\end{figure*}

Since the instantaneous part of the temporal gluon propagator
corresponds to the color-Coulomb potential,
\begin{equation}
D_{44}(\vec{x},t) = V_c(\vec{x})\delta(t) + P(\vec{x},t),
\end{equation}
the color-Coulomb potential in the color-singlet channel is given
by measuring the instantaneous $D_{44}$ in position space,
\begin{equation}
V^{\rm singlet}_c(\vec{x}) = -\frac{4}{3} a_{\tau} D_{44}(\vec{x}).
\end{equation}
Here the factor $-4/3$ is the Casimir invariant
in the fundamental representation in the SU(3) gauge group.
The instantaneous temporal gluon propagator in position space
is drawn in \Fig{fig:D44_xspace} for the isotropic and the $\xi=4$
anisotropic lattices.
We observe that the behavior of $D_{44}$ notably changes
by increasing the anisotropy.
$D_{44}$ decreases almost linearly with distance in the range $3 \le r \le 7$ [fm]
for the anisotropic lattice while it does not on the isotropic lattice.
The linear decrease of $D_{44}$ means that $V_c$ is a linearly
rising potential with distance,
which is consistent with the lattice calculations of $V_c$
from the correlator of the partial Polyakov line.
On the isotropic lattice, there are unavoidable contributions
from the polarization term in the instantaneous propagator on the lattice,
and it is indispensable to carry out the lattice simulations with
small temporal spacing to extract $V_c$ from the instantaneous $D_{44}$.


We fit the data with the function
\begin{equation}
V^{\rm singlet}_c(r) = C + \sigma_c r,
\end{equation}
in the range $0.4 \le r \le 0.7$ [fm]
and extract the color-Coulomb string tension $\sigma_c$.
The fitted result is given in \Tab{tab:fit_Vc}.
We see that the fitting is extremely worse
for the isotropic lattice, and $\chi^2/N_{\rm DF}$ approaches
$O(1)$ on the anisotropic lattice.
Besides a large $\chi^2$ value, the color-Coulomb string tension
on the isotropic lattice is smaller than the Wilson string tension
violating the Zwanziger's inequality
\cite{Zwanziger:2002sh}.
The extracted $\sigma_c$ increases with decreasing the temporal
lattice spacing and satisfies the Zwanziger's inequality
on the finer lattices.
Although $\sigma_c$ from the instantaneous $D_{44}$ is still smaller than
that from the correlator of the partial Polyakov loop,
there is little doubt that $\sigma_c$ does not saturate
the Zwanziger's inequality and $\sigma_c$ is larger than
the string tension of the Wilson static potential.

It may be surprising that we can extract $\sigma_c$
with reasonable $\chi^2$ values on finer anisotropic lattices.
We have seen in the last section that the power law fitting
of $Z_{44}$ does not work, even though the extracted value of $\gamma_{44}$
is close to 2, which gives a linearly rising color-Coulomb potential.
On the anisotropic lattice, the largest lattice volume is $4.67^4$ [fm$^4$]
(circles in the right panel of \Fig{fig:D44_xspace})
and we naively expect that the finite volume effects on the instantaneous
$D_{44}$ are not so serious in the range $r \le 2$ [fm].
However, \Fig{fig:D44_xspace} shows that $D_{44}$ does not show a linearly
decreasing behavior at $r \ge 1$ [fm].
This indicates that we still need to approach the Hamiltonian limit
to extract the color-Coulomb potential from the instantaneous temporal
gluon propagator and to compare the color-Coulomb string tension
with that from the correlator of the partial Polyakov loop.

\begin{table*}[htdp]
\caption{
The result of the fitting of $V_c$
in the range $0.4 \le r \le 0.7$ [fm].
}
\begin{center}\begin{tabular}{cccc}
\hline\hline
$( L_{\sigma}, L_{\tau}, \xi, \beta )$ & $C$ [GeV] &
$\sqrt{\sigma}$ [MeV] & $\chi^2/N_{\rm DF}$ \\ 
\hline
(48,  48, 1, 6.00) & -0.4971(4) & 289.9(3)  & 39.1 \\
(48, 192, 4, 5.75) & -1.262(2)  & 423.2(9)  & 11.3 \\
(48, 192, 4, 5.95) & -1.995(5)  & 515.7(18) & 1.26 \\
(48, 192, 4, 6.10) & -2.634(5)  & 580.0(16) & 0.828 \\
\hline\hline
\end{tabular}\end{center}
\label{tab:fit_Vc}
\end{table*}

\section{Conclusions}
\label{sec:conclusions}

It is a central interest to explore the IR behavior
of the gluon propagator to reveal the confinement mechanism
in QCD.
In the Coulomb gauge, the instantaneous gluon propagator
suffers from significant discretization errors on
isotropic lattices.
In this paper, we calculated the transverse and the temporal
components of the instantaneous gluon propagator on isotropic
and anisotropic lattices and studied the scaling behavior.

We find that the transverse gluon propagator shows a nice
scaling behavior on the anisotropic lattices,
and scaling violation observed on the isotropic lattice
almost disappear on $\xi=4$ anisotropic lattice.
It is natural to expect that the perfect scaling behavior
can be seen and the multiplicative renormalizability holds
in the Hamiltonian limit $a_{\tau}\to 0$.
In the IR region, the transverse gluon propagator is strongly
suppressed on both the isotropic and the anisotropic lattices
and shows the turnover at about 500 [MeV] in both cases.

We also calculated the transverse gluon propagator in
position space, and it shows that the correlation function
quickly decreases at small distances and becomes negative
in the range $r \sim 1-2$ [fm], and vanishes at large distances.
This means that the gluon fields have no correlation beyond
the hadronic scale, and it is consistent with the fact that
the gluons are confined in the hadrons (glueballs).

The power law fitting of the transverse gluon propagator
exhibits that the UV exponent $\gamma_{\rm gl}^{\rm UV}$ decreases
with increasing the anisotropy.
This supports the expectation that the gluons behave
as free massless particles at sufficiently large momentum
in the continuum due to asymptotic freedom.

In order to explore the IR behavior of the transverse gluon propagator,
we fitted the data with the power law and found that the extracted value
of the IR exponent is positive on the isotropic and anisotropic lattices.
However, our lattice is still small to extract a reliable value of
the IR exponent on the anisotropic lattices.
Fitting the transverse gluon propagator with the Gribov-type
ansatz did not work in our lattice data, which successfully
describe the SU(2) lattice data employing different method to
cure scaling violation
\cite{Burgio:2008jr},
and it deserves further study whether this is due to
the difference of the gauge group or the adopted prescriptions
to remedy scaling violation.

For the temporal gluon propagator, scaling violation
was observed both in the IR and UV regions on the isotropic lattice
in contrast to the transverse propagator, for which lattice data
do not suffer from discretization errors in the IR region.
Our results show that the lattice data on the anisotropic lattices
show much better scaling behavior than that on the isotropic lattice,
although the discretization errors are still seen in the IR
and the UV regions.

We observed that the time-time gluon propagator
on the anisotropic lattice is much more enhanced in the IR region
compared to that on the isotropic lattice.
The turnover observed on the coarsest lattice data on isotropic lattice
disappears on the finer lattices, and $Z_{44}$ monotonically increases
with decreasing the momentum on $\xi=4$ anisotropic lattice.
However, the IR power law fitting did not work for the dressing
function, although the fitted value $\gamma_{44}$=1.84(1)
on the finest lattice is close to the expected value
for the linearly rising behavior of the color-Coulomb potential.
We again need larger lattices as the transverse gluon propagator.

The logarithmic law fitting of the dressing function of
the temporal propagator in the UV region indicates that
the extracted $\Lambda_{\rm Coul}$ is unacceptably small
on the isotropic lattice and it takes 0.845(85) [GeV]
on the finest anisotropic lattice,
which is the order of $\Lambda_{\rm QCD}$.

In position space, the linearly decreasing behavior
can be seen on the anisotropic lattice,
and the color-Coulomb string tension obeys the Zwanziger's inequality
on finer anisotropic lattices.
However, the extracted value is still smaller than that
obtained from the correlator of the partial Polyakov line.
Thus, we can say that although the scaling violation
is softened by decreasing the temporal lattice spacing,
the instantaneous temporal gluon propagator receives a contribution
from the polarization term and it is difficult to extract
the color-Coulomb potential from the instantaneous $D_{44}$
for finite $a_{\tau}$.


\section*{Acknowledgements}

The simulation was performed on
NEC SX-8R at RCNP
and NEC SX-9 at CMC, Osaka University.
We appreciate the warm hospitality and support of the RCNP administrators.
This work is partially supported by Grant-in-Aid for JSPS
from Monbu-kagakusyo,
and 
Grant-in-Aid for Scientific Research by
Monbu-kagakusyo, Grant No. 20340055.

\begin{appendix}

\section{Tolerance of the gauge fixing and the IR suppression of $D^{\rm tr}$}
\label{sec:tolerance}

The tolerance of the gauge fixing is crucial for
the IR suppression of the instantaneous transverse gluon propagator.
In the iterative gauge fixing on the lattice,
we stop the gauge transformation
if the violation of the transversality becomes less than
some small number $\epsilon$;
\begin{equation}
\theta = \frac{1}{(N_c^2-1)L_{\sigma}^3}\sum_{\vec{x}, a, i}
(\nabla_i A_i^{\rm lat}(\vec{x},t))^2 < \epsilon.
\end{equation}
We set $\epsilon = 10^{-14}$ in our calculations.
In this appendix, we examine the behavior of the instantaneous
transverse gluon propagator by varying $\epsilon$ on the isotropic lattice.

\begin{figure}[htbp]
\begin{center}
\resizebox{0.45\textwidth}{!}
{\includegraphics{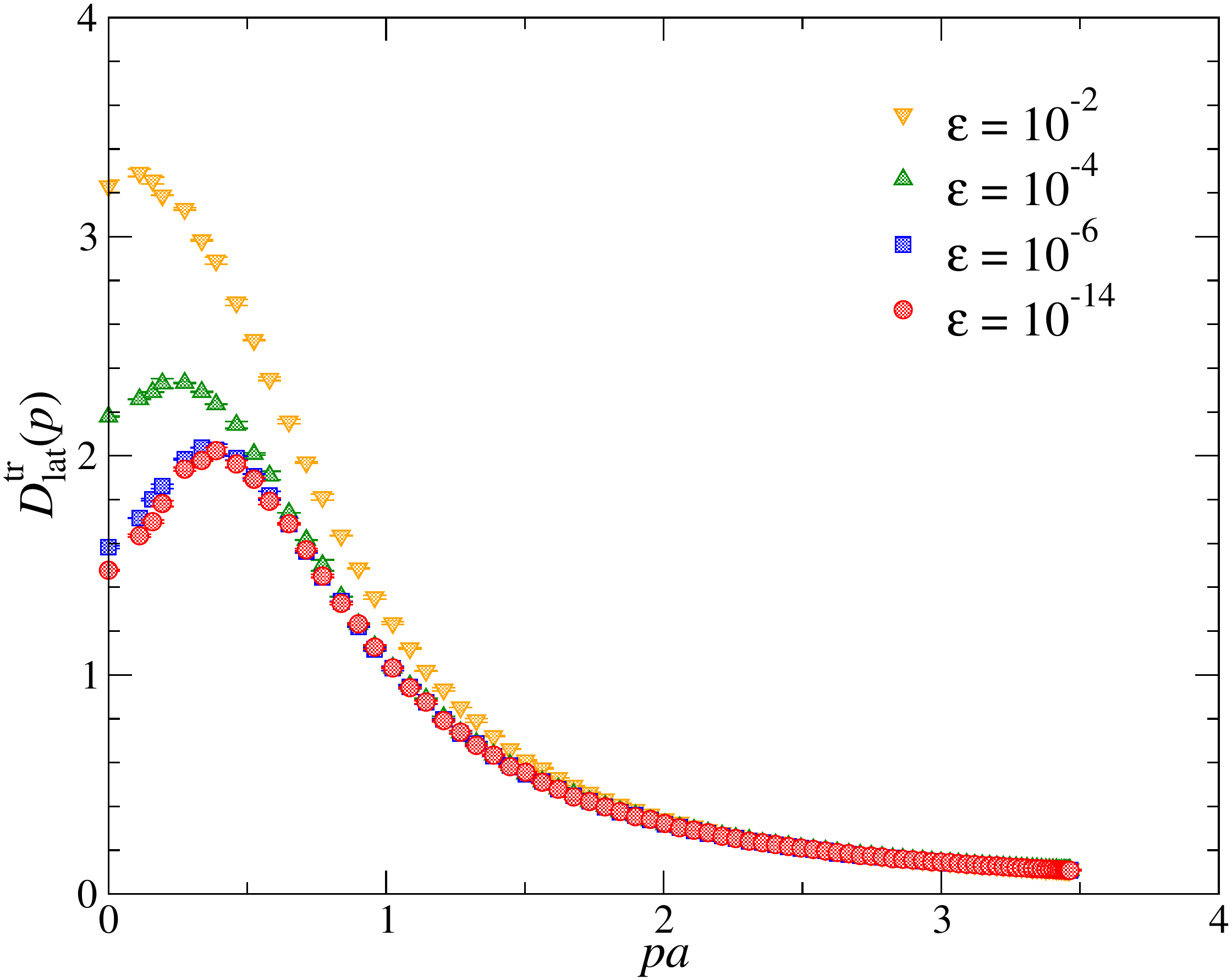}}
\end{center}
\caption{
The instantaneous transverse gluon propagator
in lattice units on $56^4$ lattice at $\beta=5.7$
for various $\epsilon$.
The cylinder cut is applied, but the cone cut is not.
}
\label{fig:Dtr_tolerance}
\end{figure}

Figure \ref{fig:Dtr_tolerance} shows $D^{\rm tr}(\vec{p})$
in lattice units on $56^4$ lattice at $\beta=5.7$ for various $\epsilon$.
For all the cases, the measurements are done for 20 configurations
and the cylinder cut is applied.
We observe that the propagator does not exhibit a turnover
for $\epsilon = 10^{-2}$ and the curve deviates from that
for $\epsilon = 10^{-14}$ in the region $pa \le 1.5$.
Decreasing $\epsilon$ by the factor 100 diminishes the propagator
at small momenta and $D^{\rm tr}$ coincides within the statistical errors
in the range $1 \le pa$.
A further decrease of $\epsilon$ suppresses $D^{\rm tr}$ in
the IR region and the transverse propagator shows a clear turnover.
Although the data for $\epsilon = 10^{-8}$ are not depicted in
\Fig{fig:Dtr_tolerance}, they fall on top of the data for
$\epsilon = 10^{-14}$ within the error bars.
Therefore, $\epsilon$ should be sufficiently small
to explore the IR behavior of the gluon propagator
and $10^{-14}$ is small enough.

\section{Scaling violation in the free field case}
\label{sec:Dtr_free}

In this appendix, we discuss scaling violation of
the instantaneous propagator in the free field case.
A more general case was discussed in
\Ref{Burgio:2008jr}.

The instantaneous transverse propagator in the continuum is given by
integrating the unequal-time propagator over $p_4$,
\begin{equation}
D(\vec{p})
= \int^{\infty}_{-\infty} \frac{dp_4}{2\pi} D(\vec{p},p_4).
\end{equation}
In the free field case (i.e., at the zeroth order in the coupling),
the four-dimensional propagator is
$D_{\rm{free}}(\vec{p},p_4) = 1/(\abs{\vec{p}}^2+p_4^2)$ and
the instantaneous propagator is
\begin{eqnarray}
D_{\rm{free}}(\vec{p})
&=& \int^{\infty}_{-\infty} \frac{dp_4}{2\pi}
\frac{1}{\abs{\vec{p}}^2+p_4^2} \nonumber \\
&=& \left. \frac{1}{2\pi\abs{\vec{p}}} \arctan\left(\frac{p_4}{\abs{\vec{p}}}\right)
\right|^{p_4=\infty}_{p_4=-\infty} \nonumber \\
&=& \frac{1}{2\abs{\vec{p}}}.
\end{eqnarray}

On a lattice with a finite temporal spacing $a_{\tau}$,
the $p_4$ integral is limited within the range
$-\pi/a_{\tau} \le p_4 \le \pi/a_{\tau}$.
Thus, the instantaneous propagator is given by
\begin{eqnarray}
D_{\rm{free}}(\vec{p},a_{\tau})
&=& \int^{\pi/a_{\tau}}_{-\pi/a_{\tau}} \frac{dp_4}{2\pi}
\frac{1}{\abs{\vec{p}}^2+p_4^2} \nonumber \\
&=& \left. \frac{1}{2\pi\abs{\vec{p}}} \arctan\left(\frac{p_4}{\abs{\vec{p}}}\right)
\right|^{p_4=\pi/a_{\tau}}_{p_4=-\pi/a_{\tau}} \nonumber \\
&=& \frac{1}{2\abs{\vec{p}}}\frac{2}{\pi}
\arctan\left(\frac{\pi}{a_{\tau}\abs{\vec{p}}}\right).
\end{eqnarray}
For a finite temporal lattice spacing,
we have an extra factor
\begin{equation}
\frac{2}{\pi}\arctan\left(\frac{\pi\xi}{\hat{p}}\right),
\end{equation}
where $\hat{p}=\abs{\vec{p}}a_{\sigma}$ is the spatial momentum in lattice units,
and $\xi$ is the anisotropy, $\xi=a_{\sigma}/a_{\tau}$.
This extra factor approaches unity in the Hamiltonian limit $a_{\tau}\to0$.

\begin{figure}[htbp]
\begin{center}
\resizebox{0.45\textwidth}{!}
{\includegraphics{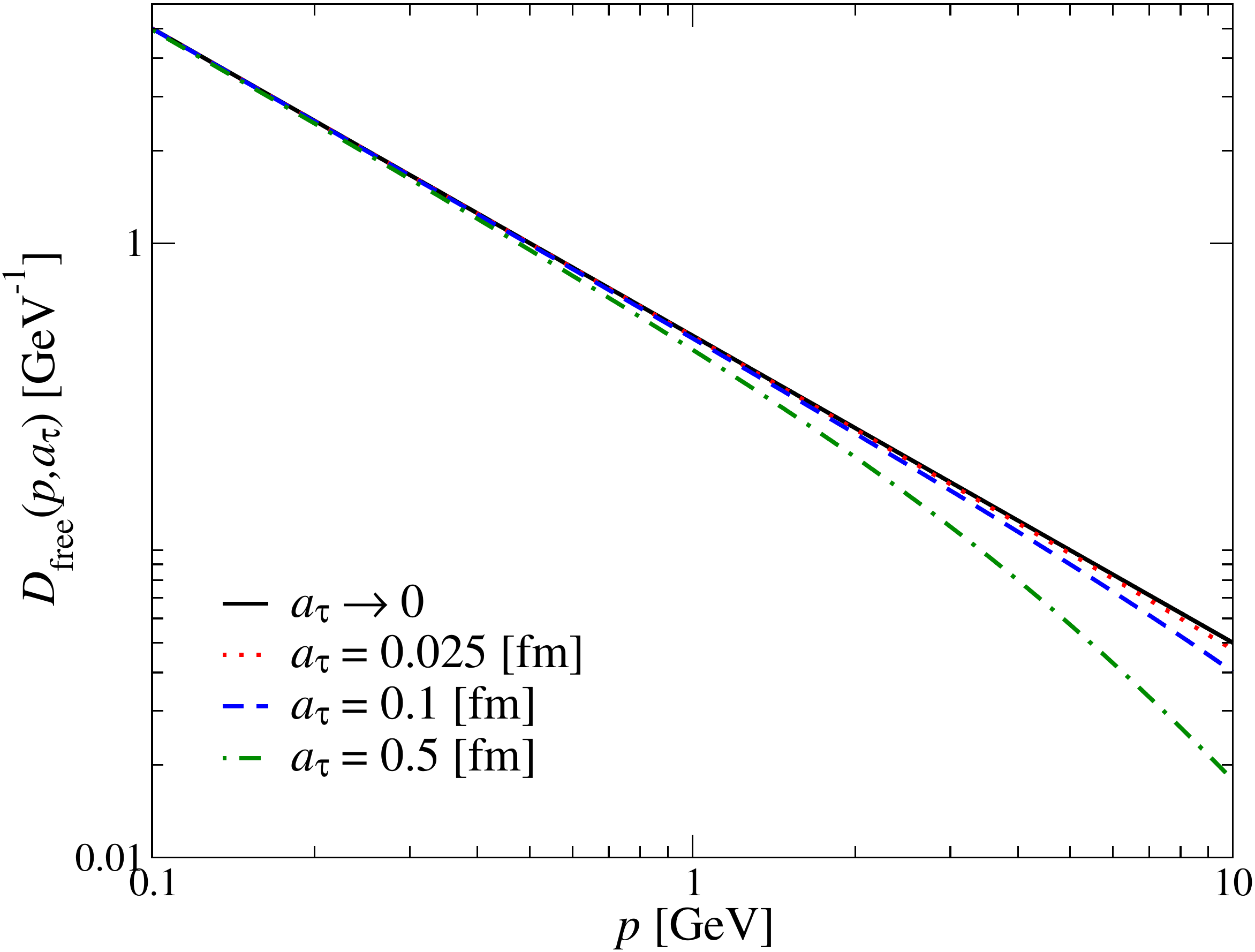}}
\end{center}
\caption{$D_{\rm{free}}(\abs{\vec{p}},a_{\tau})$ is plotted
as a function of $p$ for various temporal lattice spacings.}
\label{fig:Dtr_free}
\end{figure}

The free instantaneous propagator for various temporal lattice spacings
is illustrated in \Fig{fig:Dtr_free}.
We observe that the propagator starts to deviate from
that in the Hamiltonian limit at larger momenta as $a_{\tau}$ decreases.
Accordingly, scaling violation is observed even in the free field case
and it goes away in the Hamiltonian limit.
If we impose some renormalization condition,
$D_{\rm free}(\abs{\vec{p}}=1 {\rm [GeV]}) = 1$ [GeV$^{-1}$]
for instance, the curves in \Fig{fig:Dtr_free} cross
at the renormalization point and deviate from each other in
both the small and the large momentum regions,
which is the situation we encountered in the lattice simulations.
From this simple exercise, we expect that scaling violation
for the instantaneous gluon propagator would be moderate
for small temporal lattice spacing.

\end{appendix}


\end{document}